\documentclass[12pt,preprint]{emulateapj}

\shortauthors{Shi et al.}

\begin{document}

\title{Cosmic Evolution of Star Formation In SDSS Quasar Hosts Since $z$=1}

\author{Yong Shi\altaffilmark{1}, George H. Rieke\altaffilmark{1}, Patrick Ogle\altaffilmark{2}, Linhua Jiang\altaffilmark{1}, Aleksandar M. Diamond-Stanic\altaffilmark{1} }

\altaffiltext{1}{Steward Observatory, University of Arizona, 933 N Cherry Ave, Tucson, AZ 85721, USA}
\altaffiltext{2}{Spitzer Science Center, California Institute of Technology, Mail Code 220-6, Pasadena, CA 91125}

\begin{abstract}

We present {\it  Spitzer} IRS observations of a  complete sample of 57
SDSS type-1 quasars at $z\sim$  1. Aromatic features at 6.2 and/or 7.7
$\mu$m  are detected in  about half  of the  sample and  show profiles
similar  to   those  seen  in   normal  galaxies  at  both   low-  and
high-redshift,  indicating a  star-formation origin  for  the features.
Based  on   the  ratio  of   aromatic  to  star-formation   IR  (SFIR)
luminosities for normal star-forming galaxies at $z$ $\sim$ 1, we have
constructed the SFIR luminosity function (LF) of $z$ $\sim$ 1 quasars.
As  we found  earlier for  low-redshift  PG quasars,  these $z\sim$  1
quasars show  a flatter SFIR LF  than do $z$ $\sim$  1 field galaxies,
implying the quasar host galaxy population has on average a higher SFR
than the field galaxies do. As measured from their SFIR LF, individual
quasar hosts  have on average  LIRG-level SFRs, which mainly  arise in
the circumnuclear regions.  By  comparing with similar measurements of
low-redshift  PG quasars, we  find that  the comoving  SFIR luminosity
density in  quasar hosts  shows a much  larger increase  with redshift
than that  in field  galaxies.  The behavior  is consistent  with pure
density   evolution   since   the   average  SFR   and   the   average
SFR/BH-accretion-rate  in  quasar  hosts  show little  evolution  with
redshift.  For individual quasars, we have found a correlation between
the aromatic-based  SFR and the  luminosity of the  nuclear radiation,
consistent with  predictions of  some theoretical models.   We propose
that type 1 quasars reside  in a distinct galaxy population that shows
elliptical  morphology  but that  harbors  a  significant fraction  of
intermediate-age stars and  is experiencing intense circumnuclear star
formation.
 
\end{abstract}


\section{Introduction} 

Since its discovery, the relationship between the black-hole (BH) mass
and   galaxy    bulge   properties   \citep{Kormendy95,   Magorrian98,
  Gebhardt00,  Ferrarese00} has played  a critical  role in  our ideas
about galaxy evolution and  BH growth.  Active galactic nuclei (AGNs),
the  manifestation of accretion  onto BHs,  are the  main sites  of BH
growth.   The  demography of  local  galaxies  suggests  that most  --
perhaps   all  --  massive   galaxies  host   BHs  at   their  centers
\citep[e.g.][]{Kormendy95}.  The good match  between the local BH mass
density and the accreted BH mass density in AGNs suggests that the AGN
is an indispensable phase of galaxy evolution \citep[e.g.][]{Soltan82,
  Yu02,  Aller02, Shankar04, Marconi04}.   The interplay  between star
formation and AGN activity may play a key role in the establishment of
the  BH-bulge  correlation.   However,  in  quasars  (luminous  AGNs),
observational  constrains on  the  BH/star-formation relationship  are
rare due to the difficulty  in measuring the star formation rate (SFR)
around the bright quasar nuclei.

The host galaxy morphology and  colors provide the first insight into
the level of star formation.   At low redshift ($z<$ 0.5), decades of
effort   have  been  devoted   to  imaging   the  host   galaxies  of
UV/optically-selected  type  1  quasars \citep{Hutchings84,  Smith86,
  McLeod95,   Bahcall97,  Dunlop03,  Floyd04,   Guyon06}.   Important
technical  developments,  especially  with space-based  and  adaptive
optics (AO) instruments, have provided observations with dramatically
improved  spatial resolution,  stable PSFs  and large  dynamic range,
which are  keys to extracting quantitative information  on host
galaxies surrounding the bright AGN.   Radio-loud quasars are found to
be  associated   with  luminous  early-type   galaxies  with  structure
parameters  indistinguishable  from  those of  quiescent  ellipticals
\citep{Smith86, Dunlop03}.  Radio-quiet  quasars reside on average in
less  luminous  galaxies with  diverse  morphologies  that show  some
dependence on the host  and nuclear luminosity, i.e., more frequently
of  early  type  for the more   luminous  nuclei  and  hosts  \citep{McLeod95,
  Bahcall97, Floyd04, Guyon06}.  At  $M_{B}$ $<$ -23, roughly 50\% of
quasar  hosts show  normal elliptical  morphologies \citep{Bahcall97,
  Guyon06}.  The fraction of bright quasar hosts with ongoing merging
features is 10-30\% \citep{Bahcall97, Guyon06}.  However, this merger
fraction may be under-estimated  especially at the late merging stage
where the tidal tails become  relatively faint.  Such a bias has been
demonstrated  by  the discoveries  of  merger  remnants in  quiescent
elliptical hosts \citep{Canalizo07,  Bennert08}.  The merger fraction
for  near-IR-selected  quasars   \citep{Cutri02}  and  far-IR  excess
quasars     \citep{Canalizo01}    is     much     higher    ($>$50\%)
\citep{Canalizo01, Hutchings03, Marble03}.  However, it is unclear if
this  is because  tidal features  are relatively  easily  detected in
IR-selected quasars with low  nuclear/host-light contrast or if there
is  an  evolutionary  difference  between  quasars  selected  in  the
UV/optical and those selected in the IR.

The  study of  stellar populations  in quasar  hosts  using UV/optical
spectra  has provided  another  significant constraint  on their  star
formation  properties, mainly  through detecting  the presence  of the
young-  ($\lesssim$100 Myr),  intermediate-  ($\sim$0.1-1Gyr) or  old-
($\sim$10Gyr) stellar populations.  However, in both type-1 and type-2
quasars,  the nuclear  radiation  strongly dilutes  the young  stellar
features, such as  the PCygni UV lines and the  4650$\AA$ WR bump, and
dominates the line  emission of SFR tracers used  for normal galaxies.
The mean-age-indicator  4000$\AA$ break and intermediate-age-indicator
($\sim$0.1-1 Gyr) H$\delta$ absorption  line also suffer from dilution
by nuclear  light.  Despite these difficulties,  and inconsistent with
the   implication   of    the   early-type   host   morphologies,   an
intermediate-age stellar population is present at a significant level.
Off-nuclear ($>$$\sim$15  kpc) UV/optical spectra  of quasars indicate
that    old     stellar    populations    dominate    \citep{Hughes00,
  Nolan01}. However,  with much deeper spectra  a significant fraction
(10\% in mass fraction)  of an intermediate-age population is revealed
\citep{Canalizo06,  Canalizo09}.   In type-2  AGNs  and less  luminous
type-1  objects  where the  nuclear  radiation  contamination is  less
severe,  the presence  of an  intermediate-age stellar  population has
been    extensively    confirmed   \citep{Kotilainen94,    Ronnback96,
  Brotherton99,   Kauffmann03,   Jahnke04,  VandenBerk06,   Jahnke07}.
Furthermore, it seems there is a decrease in the mean stellar age with
increasing AGN luminosity \citep{Kauffmann03, VandenBerk06}.

Techniques  that  are   more  suitable  for  detecting  ongoing/recent
($\lesssim$  100  Myr) star  formation  have  been employed  recently,
although their  accuracy is still  not comparable to those  for normal
galaxies.   As the  material  reservoir of  star  formation, the  cold
molecular  gas  mass  roughly   correlates  with  the  level  of  star
formation.   It has  been found  that quasar  hosts are  rich  in cold
molecular   gas  \citep{Scoville03,   Evans06,   Bertram07},  implying
significant ongoing star formation  activity.  With space-based and AO
instruments,  spatially-resolved images of  SFR tracers  are promising
ways to quantify the SFR  and characterize the spatial distribution of
the  star-forming  regions.   For  example,  the  extended  Pa$\alpha$
emission  in PG1126-041  is  most likely  associated  with an  intense
nuclear (100pc) starburst embedded  in the old bulge \citep{Cresci04}.

A new method  to measure the SFR  of local quasar hosts is  to use the
mid-infrared aromatic bands  \citep{Ogle06, Schweitzer06, Shi07, Fu09}
measured with  the Spitzer Space Telescope.  In  field galaxies, these
emission  bands are  universally bright  in roughly  solar metallicity
regions of  star formation, with well understood  relationships to the
SFR   \citep[e.g.][]{Roussel01,  Dale02,   Wu05,   Brandl06,  Smith07,
  Engelbracht08}.  Their carriers appear  to be destroyed by the harsh
radiation fields around high-luminosity AGNs \citep{Genzel98, Spoon07}
and  thus any  detected  aromatic  emission is  likely  from the  host
galaxy.

In \citet{Shi07}, we provided  further evidence for the star-formation
origin of the aromatic features  in quasar hosts: (1) both the overall
shape  of   the  aromatic  features   and  the  distribution   of  the
7.7$\mu$m/11.3$\mu$m  feature  ratio  are  similar to  those  in  star
forming galaxies; (2) the larger the equivalent widths of the aromatic
features,  the stronger  the far-IR  emission  in the  global SEDs  of
quasars; (3) quasars lie on the  trend of normal galaxies in the plane
of the molecular  gas mass and the aromatic-derived  SF IR luminosity.
By  measuring the  aromatic feature  fluxes in  a large  quasar sample
($\sim$200  objects  including  PG,  2MASS  \&  3CR),  we  obtained  a
quantitative census of star  formation activity in low-redshift quasar
hosts \citep{Shi07}.  We showed that  quasar hosts have a flatter star
formation  infrared   (SFIR)  luminosity  function   (LF)  than  field
galaxies, i.e., the quasars lie in very luminous star forming galaxies
more often  than would be the  case for a random  sampling of galaxies
without nuclear activity.

As a summary, type-1 quasars  at low redshift ($z\lesssim$ 0.5) mainly
reside in  luminous early-type host galaxies but  harbor a significant
($>$10\%)  fraction of  intermediate-age stellar  populations  and are
experiencing  intense star formation  activity.  These  properties are
rare  for normal  galaxies,  implying  that the  quasar  appears at  a
special stage  of galaxy evolution.   In this paper, we  probe whether
type-1   quasar  host   galaxies  at   $z$  $\sim$   1   have  similar
characteristics.

As demonstrated  in our low-redshift work  \citep{Shi07}, the aromatic
feature observed with {\it Spitzer} provides efficient measurements of
the  SFR in  quasar hosts  with significantly  improved  accuracy.  To
constrain the star-formation/quasar interplay  at high redshift and to
characterize the  cosmic evolution of star formation  in quasar hosts,
we have carried out  IRS observations of a complete optically-selected
type-1  quasar sample  (57 objects)  at $z\sim$  1 to  constrain their
SFRs.   The  sample selection  and  data  reduction  are presented  in
\S~\ref{sample-data}.  The measurement of the aromatic feature flux is
discussed in \S~\ref{MEA-PAH}.  We show the results in \S~\ref{result}
and discuss them in  \S~\ref{discussion}.  The conclusion is presented
in \S~\ref{conclusion}.   Throughout the  paper, we adopt  a cosmology
with  $H_{0}$=70  km  s$^{-1}$  Mpc$^{-1}$, $\Omega_{\rm  m}$=0.3  and
$\Omega_{\Lambda}$=0.7. All magnitudes are defined in the AB magnitude
system.

\section{Sample and Data Reduction}\label{sample-data}

\begin{deluxetable*}{llllcllll}
\tablecolumns{8}
\tabletypesize{\scriptsize}
\tablewidth{0pc}
\tablecaption{\label{PAH_Table} Aromatic Features in $z\sim$1 Quasar Hosts}
\tablehead{
  \colhead{Sources}         & \colhead{z}                 & \colhead{$m_{i}$}           & 
  \colhead{$M_{\rm g}$}      & \colhead{$F_{6.2{\rm PAH}}$} & \colhead{$F_{7.7{\rm PAH}}$} &
  \colhead{$L_{\rm TIR}^{\rm SF}$}    & \colhead{$L_{5{\textendash}6{\mu}m}$}  \\
  \colhead{   }            & \colhead{   }              &   \colhead{}                 & 
  \colhead{}               & \colhead{[$10^{-15}$erg s$^{-1}$ cm$^{-2}$]}               &   
  \colhead{[$10^{-15}$erg s$^{-1}$ cm$^{-2}$]}           & \colhead{[$10^{11}$L$_{\odot}$]}
  & \colhead{[$10^{11}$L$_{\odot}$]}   \\
  \colhead{(1)}            & \colhead{(2)}            &  \colhead{(3)}                 &   
  \colhead{(4)}            & \colhead{(5)}            &  \colhead{(6)}                 &
  \colhead{(7)}            & \colhead{(8)} }
\startdata
SDSS103855.33+575814.7 &  0.96 & 18.56 & -24.87 & $<$ 2.0      &               & $<$ 3.07           &  1.19 \\
SDSS104239.66+583231.0 &  1.00 & 19.06 & -24.44 & $<$ 2.2      &               & $<$ 4.24           &  0.44 \\
SDSS105000.21+581904.2 &  0.83 & 17.76 & -25.38 & $<$ 1.5      &               & $<$ 1.19           &  1.00 \\
SDSS154542.78+505759.2 &  0.94 & 19.04 & -24.35 & $<$ 1.4      &               & $<$ 1.67           &  0.52 \\
SDSS155853.64+425817.7 &  0.87 & 18.50 & -24.75 &  1.5$\pm$0.5 &               &     1.43$\pm$ 1.17 &  0.51 \\
SDSS155934.87+380741.4 &  0.96 & 18.39 & -25.03 & $<$ 1.6      &               & $<$ 2.25           &  0.54 \\
SDSS160013.44+383406.2 &  0.95 & 19.02 & -24.39 &  3.6$\pm$0.5 &               &     7.12$\pm$ 4.79 &  0.78 \\
SDSS160318.60+445349.0 &  0.98 & 19.05 & -24.42 & $<$ 2.0      &               & $<$ 3.45           &  0.71 \\
SDSS160441.85+463756.8 &  0.94 & 19.09 & -24.30 & $<$ 1.3      &               & $<$ 1.56           &  0.40 \\
SDSS160527.54+372801.9 &  0.82 & 17.57 & -25.53 & $<$ 1.8      &               & $<$ 1.47           &  1.12 \\
SDSS160659.59+414919.3 &  0.98 & 18.97 & -24.49 & $<$ 1.4      &               & $<$ 2.10           &  0.57 \\
SDSS160733.26+451610.6 &  0.94 & 18.93 & -24.47 &  2.9$\pm$0.5 &               &     5.04$\pm$ 3.49 &  0.54 \\
SDSS160738.49+465753.6 &  0.97 & 19.06 & -24.38 & $<$ 1.4      &               & $<$ 1.94           &  0.97 \\
SDSS160822.52+415852.7 &  0.84 & 17.42 & -25.74 & $<$ 1.4      &  10.9$\pm$2.0 &     2.92$\pm$ 2.16 &  1.78 \\
SDSS160855.43+435259.2 &  0.93 & 17.25 & -26.13 & $<$ 1.7      &               & $<$ 2.25           &  4.23 \\
SDSS160904.64+405547.5 &  0.89 & 18.70 & -24.58 &  3.9$\pm$0.4 &  16.0$\pm$4.0 &     6.18$\pm$ 4.21 &  0.89 \\
SDSS161055.65+380305.7 &  0.83 & 18.54 & -24.60 &  4.1$\pm$0.4 &  15.5$\pm$1.8 &     5.26$\pm$ 3.63 &  0.66 \\
SDSS161104.75+462052.6 &  0.83 & 18.80 & -24.33 & $<$ 1.4      &               & $<$ 1.06           &  1.82 \\
SDSS161144.01+370330.7 &  0.81 & 17.77 & -25.31 &  2.7$\pm$0.6 &  11.2$\pm$1.7 &     2.65$\pm$ 1.99 &  1.15 \\
SDSS161252.50+403032.0 &  0.82 & 17.16 & -25.96 &  6.2$\pm$0.5 &  28.3$\pm$1.8 &     9.16$\pm$ 6.07 &  3.00 \\
SDSS161342.97+390732.8 &  0.98 & 18.25 & -25.21 & $<$ 1.5      &               & $<$ 2.30           &  1.07 \\
SDSS161351.34+374258.7 &  0.81 & 18.87 & -24.21 & $<$ 1.6      &               & $<$ 1.23           &  0.42 \\
SDSS161456.56+460744.0 &  0.85 & 18.79 & -24.38 &  2.2$\pm$0.5 &  21.2$\pm$2.3 &     2.28$\pm$ 1.75 &  1.14 \\
SDSS161637.16+390356.8 &  0.81 & 17.89 & -25.19 &  3.2$\pm$0.8 &  14.0$\pm$1.7 &     3.42$\pm$ 2.48 &  1.72 \\
SDSS161756.77+423924.1 &  0.97 & 18.62 & -24.84 &  2.3$\pm$0.5 &               &     4.08$\pm$ 2.89 &  0.80 \\
SDSS161806.31+422532.1 &  0.93 & 17.94 & -25.44 & $<$ 1.6      &               & $<$ 2.01           &  2.40 \\
SDSS161946.29+435915.4 &  0.85 & 18.61 & -24.58 &  2.0$\pm$0.6 &               &     2.01$\pm$ 1.57 &  0.40 \\
SDSS162035.63+420742.6 &  0.96 & 19.08 & -24.36 & $<$ 1.9      &               & $<$ 2.93           &  0.48 \\
SDSS162051.17+423449.3 &  0.99 & 18.88 & -24.61 & $<$ 1.7      &               & $<$ 2.79           &  1.69 \\
SDSS162058.08+420424.5 &  0.96 & 18.84 & -24.60 & $<$ 1.9      &               & $<$ 2.93           &  0.39 \\
SDSS162110.33+361358.6 &  0.83 & 17.88 & -25.26 & $<$ 1.3      &               & $<$ 0.99           &  0.79 \\
SDSS162123.86+425229.4 &  0.98 & 18.42 & -25.04 &  3.8$\pm$0.5 &               &     8.44$\pm$ 5.61 &  1.27 \\
SDSS162135.65+395452.1 &  0.98 & 19.05 & -24.41 & $<$ 1.8      &               & $<$ 2.87           &  0.50 \\
SDSS162224.66+371300.9 &  0.82 & 19.05 & -24.05 &  2.8$\pm$0.4 &  10.6$\pm$1.7 &     2.88$\pm$ 2.13 &  0.48 \\
SDSS162248.32+403029.4 &  0.81 & 18.93 & -24.14 & $<$ 1.6      &               & $<$ 1.23           &  0.64 \\
SDSS162318.89+402258.7 &  0.91 & 17.78 & -25.55 & $<$ 1.4      &               & $<$ 1.50           &  1.20 \\
SDSS162330.53+355933.2 &  0.87 & 18.46 & -24.76 & $<$ 1.5      &  11.8$\pm$3.1 &     3.72$\pm$ 2.67 &  1.15 \\
SDSS162349.46+364721.7 &  0.92 & 18.49 & -24.87 & $<$ 1.6      &               & $<$ 1.93           &  0.72 \\
SDSS162449.96+350857.9 &  0.88 & 18.94 & -24.32 & $<$ 1.6      &               & $<$ 1.69           &  0.68 \\
SDSS162658.11+353030.1 &  0.85 & 18.76 & -24.43 & $<$ 1.5      &   7.7$\pm$2.1 &     1.84$\pm$ 1.46 &  0.55 \\
SDSS162902.59+372430.8 &  0.93 & 19.03 & -24.34 &  3.2$\pm$0.5 &               &     5.60$\pm$ 3.84 &  0.76 \\
SDSS163018.71+371904.5 &  0.97 & 18.39 & -25.06 & $<$ 1.5      &               & $<$ 2.04           &  1.84 \\
SDSS163225.56+411852.4 &  0.91 & 18.42 & -24.91 &  1.7$\pm$0.4 &               &     2.07$\pm$ 1.61 &  0.83 \\
SDSS163537.99+365936.6 &  0.90 & 18.35 & -24.97 & $<$ 1.4      &               & $<$ 1.53           &  1.26 \\
SDSS163624.98+361457.9 &  0.91 & 18.93 & -24.40 & $<$ 1.5      &  17.8$\pm$4.5 &     7.92$\pm$ 5.29 &  0.33 \\
SDSS163656.84+364340.4 &  0.85 & 18.54 & -24.65 & $<$ 1.5      &   7.4$\pm$2.4 &     1.72$\pm$ 1.38 &  1.19 \\
SDSS163726.88+404432.9 &  0.86 & 18.52 & -24.68 &  4.1$\pm$0.4 &   7.9$\pm$2.3 &     6.03$\pm$ 4.11 &  0.66 \\
SDSS163926.19+390821.4 &  0.88 & 18.88 & -24.37 & $<$ 1.4      &               & $<$ 1.30           &  0.21 \\
SDSS164334.80+390102.5 &  0.93 & 18.91 & -24.45 & $<$ 1.6      &               & $<$ 1.97           &  0.54 \\
SDSS164346.66+383812.9 &  0.89 & 18.94 & -24.35 &  5.6$\pm$0.4 &  13.6$\pm$3.6 &    10.50$\pm$ 6.92 &  0.36 \\
SDSS164455.12+384936.3 &  0.86 & 18.78 & -24.43 & $<$ 1.5      &  13.5$\pm$3.1 &     4.34$\pm$ 3.06 &  1.23 \\
SDSS164508.74+374057.0 &  0.93 & 19.02 & -24.35 &  1.9$\pm$0.5 &               &     2.58$\pm$ 1.94 &  0.73 \\
SDSS164600.09+381833.3 &  0.99 & 18.18 & -25.30 & $<$ 1.5      &               & $<$ 2.36           &  2.20 \\
SDSS164730.89+360101.6 &  0.85 & 18.78 & -24.40 & $<$ 1.4      &   7.1$\pm$2.3 &     1.64$\pm$ 1.32 &  0.63 \\
SDSS164745.05+355732.9 &  0.94 & 19.07 & -24.33 &  1.4$\pm$0.5 &               &     1.81$\pm$ 1.44 &  0.43 \\
SDSS164925.07+373015.4 &  0.95 & 18.97 & -24.44 &  4.9$\pm$0.5 &               &    10.96$\pm$ 7.21 &  0.53 \\
SDSS165614.00+351014.5 &  0.81 & 17.44 & -25.63 &  3.8$\pm$0.6 &  29.9$\pm$1.7 &     4.37$\pm$ 3.08 &  1.73 \\
LogL$_{5-6{\mu}m}$ $\in$ [  8.0, 10.8] &  & &      &            &             & $ 0.18^{+0.09}_{-0.12}$ & $ 0.43^{+0.14}_{-0.23}$ \\
LogL$_{5-6{\mu}m}$ $\in$ [ 10.8, 12.5] &  & &      &            &             & $ 0.25^{+1.22}_{-0.15}$ & $ 1.24^{+2.99}_{-0.60}$ \\
\enddata
\tablecomments{ Col.(1): Source name. Col.(2): Redshift. Col.(3): the i-band magnitude. Col.(4):
the rest-frame g-band magnitude. Col.(5): the 6.2 $\mu$m aromatic feature flux and 3$\sigma$ upperlimits. 
Col.(6): the 7.7 $\mu$m aromatic feature flux. Col.(7): the PAH-derived
total star-forming IR luminosity. Col.(8): the continuum luminosity integrated from 5 to 6 $\mu$m.\\
The last two rows are for stacked spectra of individually PAH-undetected objects within two
5-6 $\mu$m luminosity ranges.}
\end{deluxetable*}

\setcounter{figure}{0}
\begin{figure*}
\epsscale{.90}
\plotone{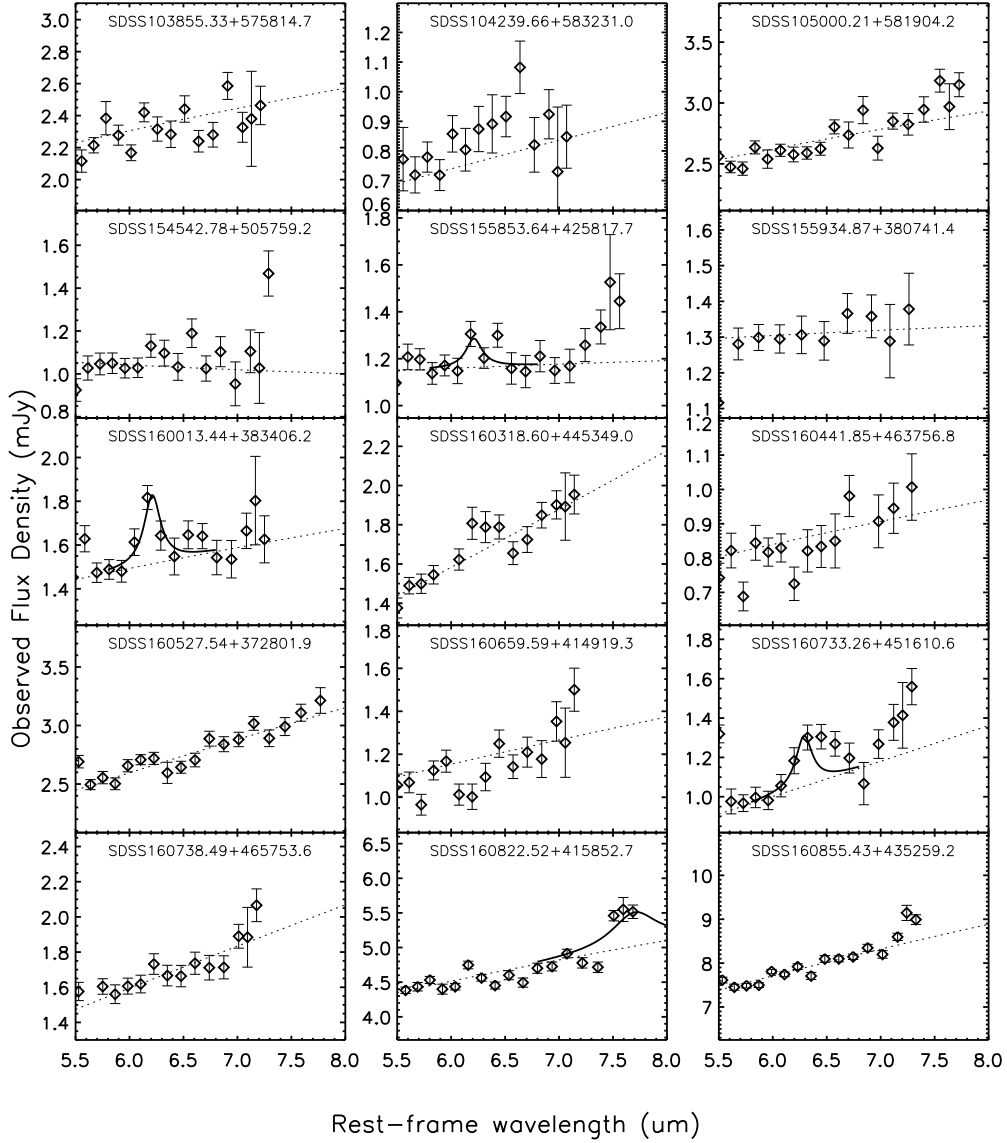}
\caption{\label{all_spec} The IRS spectra in the rest-frame wavelength ($\mu$m)
vs. the observed  flux density (mJy). The dotted lines  show the power-law
continua  while the  solid lines  indicate the  fitted 6.2  and 7.7
$\mu$m features.}
\end{figure*}

\setcounter{figure}{0}
\begin{figure*}
\epsscale{1.0}
\plotone{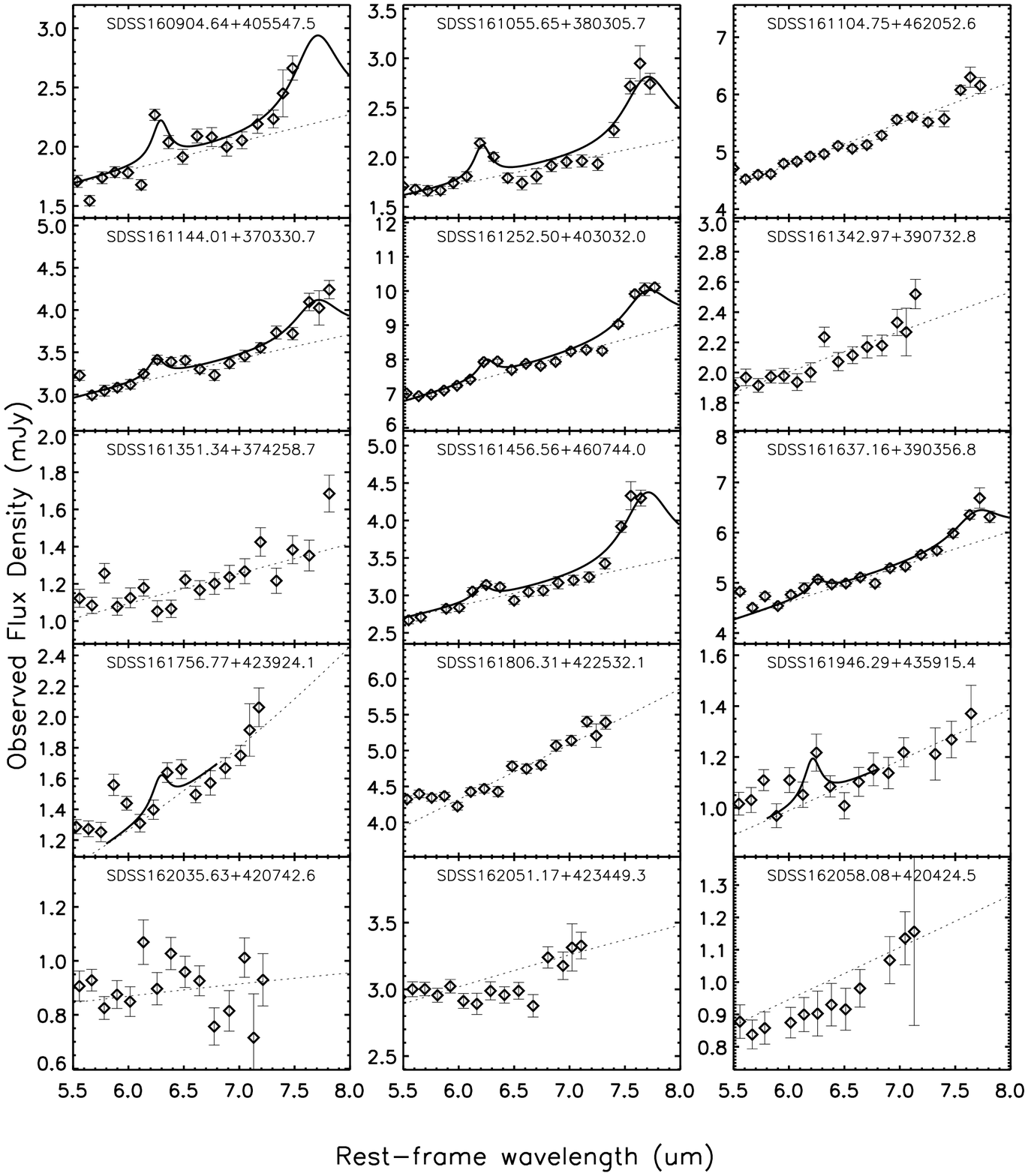}
\caption{Continued}
\end{figure*}

\setcounter{figure}{0}
\begin{figure*}
\epsscale{1.0}
\plotone{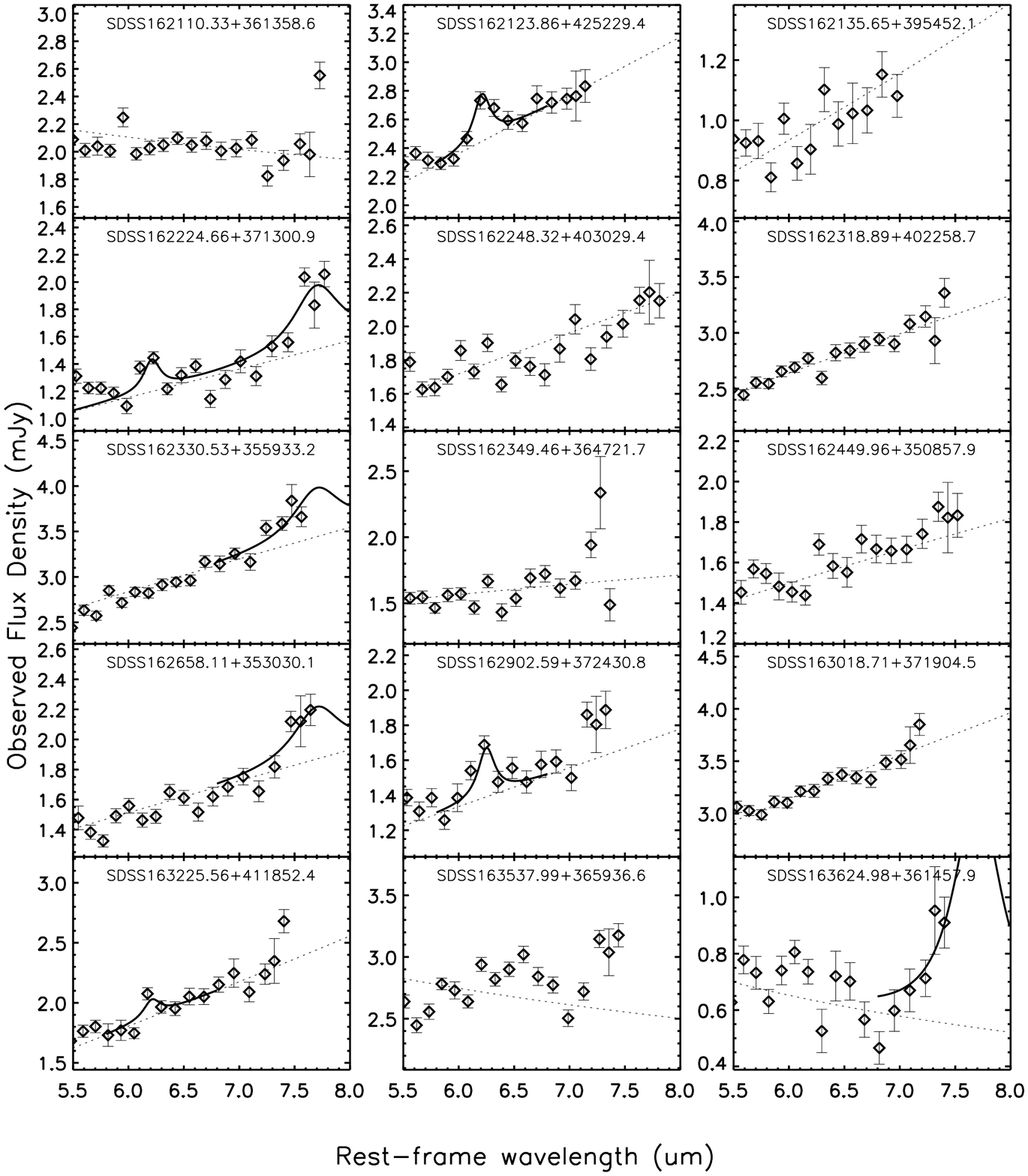}
\caption{Continued}
\end{figure*}

\setcounter{figure}{0}
\begin{figure*}
\epsscale{1.0}
\plotone{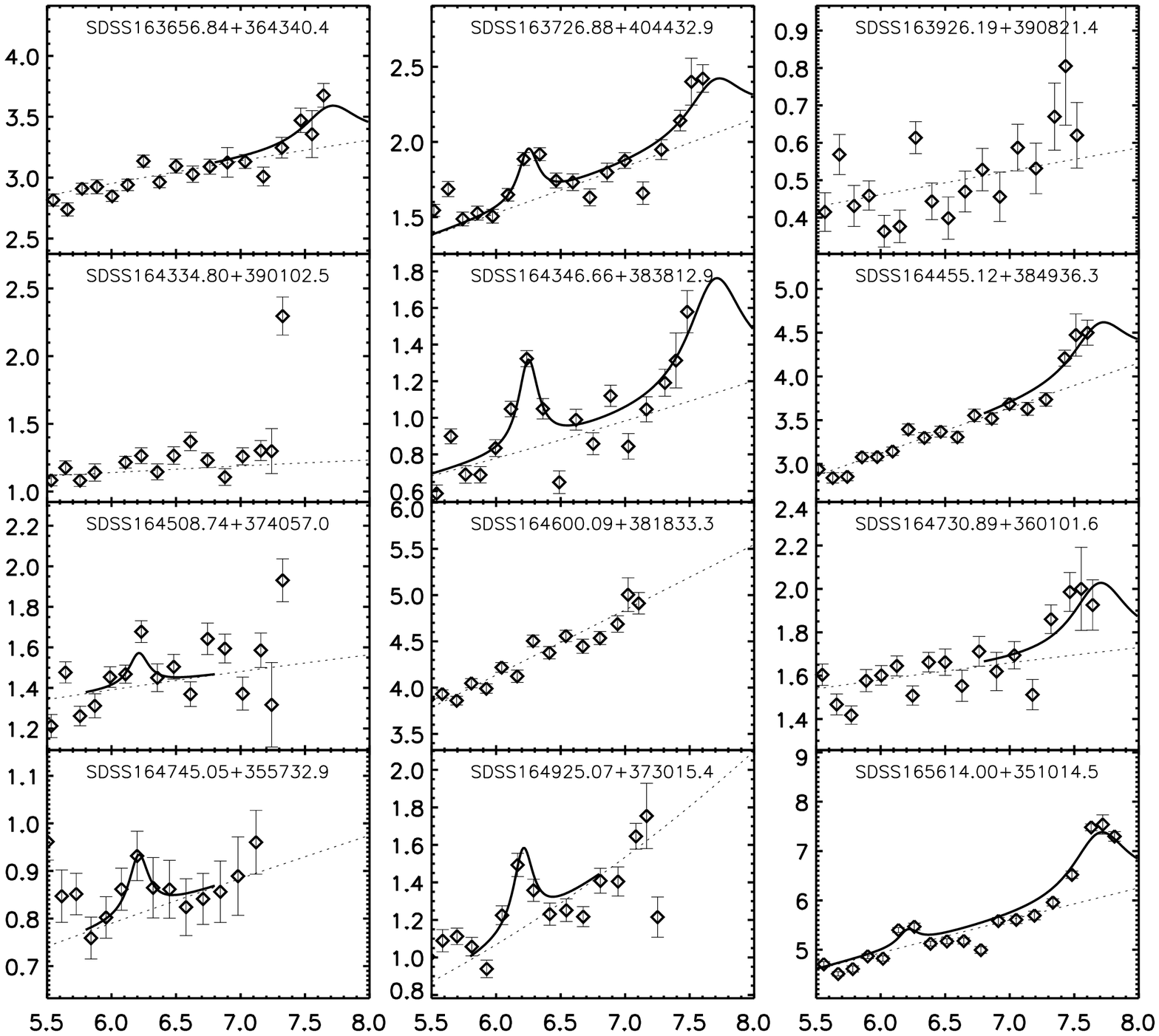}
\caption{Continued}
\end{figure*}


Our quasar sample (Table~\ref{PAH_Table}) is a  subset of the nearly homogeneous parent sample
from \citet{Richards06}.  The completeness  for this parent sample has
been well quantified and it has  been used to construct the quasar LF.
To observe  objects in low IR  background, we selected  all targets at
0.8 $\leq$ $z$ $\leq$ 1.0  from the \citet{Richards06} sample in three
fields centered  at but significantly larger than  three {\it Spitzer}
deep   fields,  including   the  {\it   Spitzer}   Wide-Area  Infrared
Extragalactic  Survey  (SWIRE)  ELAIS-N1,  SWIRE  ELAIS-N2  and  SWIRE
Lockman     Hole.     Our     three    fields     have     sizes    of
15.85$^{\circ}$$\times$9.01$^{\circ}$     centered     at    16h11m00s
+55$^{\circ}$00$''$00$'$,         19.22$^{\circ}$$\times$12.0$^{\circ}$
centered      at      16h36m48s      41$^{\circ}$01$''$45$'$,      and
3.77$^{\circ}$$\times$3.0$^{\circ}$      centered     at     10h45m00s
+58$^{\circ}$00$'$00$''$, respectively.  Our  final sample is composed
of 57 SDSS  quasars at 0.8 $\leq$ $z$ $\leq$ 1.0  and is complete down
to SDSS  i=19.0. As the SDSS  quasar coverage does not  fill the above
three  fields, the  effective area  is calculated  by  subdividing the
field into small  rectangles (about 1 square degree)  and then summing
those falling within the actual survey region. This gives an effective
area of  88.0 square degree, which is  almost the same as  the value (83.0
square degree) derived  from the total effective area  of the SDSS-DR3
survey and the  relative number of SDSS-DR3 quasars  at 0.8 $\leq$ $z$
$\leq$ 1.0  included in our  sample.  Although we observed  quasars in
low IR  background, there  are no selection  criteria based on  the IR
fluxes of individual objects.

We  obtained IRS  spectra of  the  members of  this sample  (PI-George
Rieke, PID-50196). For  objects in the redshift range  between 0.8 and
1.0, IRS  SL1 (7.4-14.5$\mu$m) covers the 6.2  $\mu$m aromatic feature
and  the  spectral  range  required for  continuum  subtraction.   For
objects at $z$  $<$ 0.9, the blue wing of the  7.7 $\mu$m feature will
also be observed if it is  present.  The IRS staring mode was used for
the  observations   and  the  exposure   time  for  each   object  was
240$\times$2 secs.

The data reduction basically followed the IRS data handbook.  Briefly,
each BCD image was first  cleaned using IRSCLEAN.  The images observed
on the  same slit position  were sigma-clipped, averaged, and  used to
subtract the  sky for  each image observed  on another  slit position.
The spectra were extracted from each sky-subtracted BCD image with the
optimal             extraction             algorithm             using
SPICE \footnote{http://ssc.spitzer.caltech.edu/postbcd/spice.html} and
then  sigma-clipped  and  averaged   to  produce  the  final  combined
spectra. We individually inspected each flux profile along the slit to
make sure the extraction aperture was centered on the brightness peak.

\section{Measurements of Aromatic Features}\label{MEA-PAH}

Because we can work at shorter rest wavelengths, the extraction of the
aromatic features  is relatively straightforward compared  to that for
local  AGNs in \citet{Shi07}.   In that  case, careful  subtraction of
silicate  features was  needed  to  measure the  7.7  and 11.3  $\mu$m
aromatic  bands  and  Monte-Carlo  simulations  were  carried  out  to
characterize the measurement  uncertainties. In the current situation,
for each  spectrum the continuum is  fitted with a power  law shape to
the flux density  in the spectral range of  5.8-6.0 $\mu$m and 6.8-7.1
$\mu$m. We then fitted the continuum-subtracted spectra with two Drude
profiles, one centered at 6.22 $\mu$m  with a FWHM of 0.186 $\mu$m and
another centered at 7.7 $\mu$m with  a FWHM of 0.53 $\mu$m for the 6.2
$\mu$m and  7.7 $\mu$m features,  respectively \citep[see][]{Smith07}.
For several objects, the central  wavelength of the 6.2 $\mu$m feature
is adjusted between 6.2 and 6.3 $\mu$m to achieve the best fit.

The feature flux was obtained by integrating the fitted profile from 1
to 20 $\mu$m.  For the 7.7  $\mu$m feature, the flux was only measured
if the rest-frame spectrum  wavelength extends longer than 7.4 $\mu$m.
The noise  of the 6.2 $\mu$m  feature was defined  within the spectral
range between 6.0 $\mu$m and  6.4 $\mu$m.  For the 7.7 $\mu$m feature,
the  S/N is  defined in  the  spectral range  from 7.0  $\mu$m to  the
longest wavelength (usually $<$ 8$\mu$m).  Aromatic features with S/Ns
above 3  were considered  to be solid  detections.  We  inspected each
detected feature  and excluded several  detections due to  large noise
around the feature. The final detection rate is $\sim$37\% for the 6.2
$\mu$m feature  and $\sim$50\%  for any feature  at either 6.2  or 7.7
$\mu$m.    The   spectra   and    fitted   features   are   shown   in
Figure~\ref{all_spec}. The fluxes or 3$\sigma$ upper-limits are listed
in Table~\ref{PAH_Table}.

For the  7.7 $\mu$m feature,  a test based  on the four  nearby galaxy
composite spectra of  \citet{Smith07} shows that a fit  only using the
blue wing does  not introduce a large uncertainty.  On the other hand,
the silicate emission feature  and sometimes [NeV]7.65 $\mu$m line can
be  present in  these type  1 quasars  \citep{Shi06}. To  quantify the
effect of  these two  features on the  7.7 $\mu$m aromatic  fluxes, we
have used  the above method  to measure the  feature fluxes in  all PG
quasars.  They  are  compared   to  more  accurate  ones  obtained  in
\citet{Shi07}     by     deriving    ($f_{\rm     7.7{\mu}mPAH}-f_{\rm
  7.7{\mu}mPAH}^{0})/f_{\rm    7.7{\mu}mPAH}^{0}$,    where    $f_{\rm
  7.7{\mu}mPAH}$ is  the flux  based on the  method of this  paper and
$f_{\rm 7.7{\mu}mPAH}^{0}$  is the one derived  in \citet{Shi07}. This
ratio has  an average  value of  -0.2 and a  68\% confidence  range of
-0.55 to  0.15.  The differences are  mainly caused by  the deviation of
the IR spectrum at $\lambda$ $>$ 8 $\mu$m from the interpolation based
on the part below 8 $\mu$m. Such a deviation is due to either a change
in the IR continuum or  the presence of the silicate emission feature.
Ten objects out  of fifty-seven have both 6.2  and 7.7 $\mu$m features
and seven  more have detections only  of the 7.7  $\mu$m feature.  The
median  value  of  the  $f_{\rm  7.7PAH}$/$f_{\rm  6.2PAH}$  ratio  is
4.1$\pm$2.4. This  ratio is used  to obtain the equivalent  6.2 $\mu$m
feature flux for those objects with only 7.7 $\mu$m fluxes.

\section{RESULTS}\label{result}

\subsection{IR Spectra of $z$$\sim$1 Quasars And Comparisons To Other Samples}

\begin{figure*}
\epsscale{1.}
\plotone{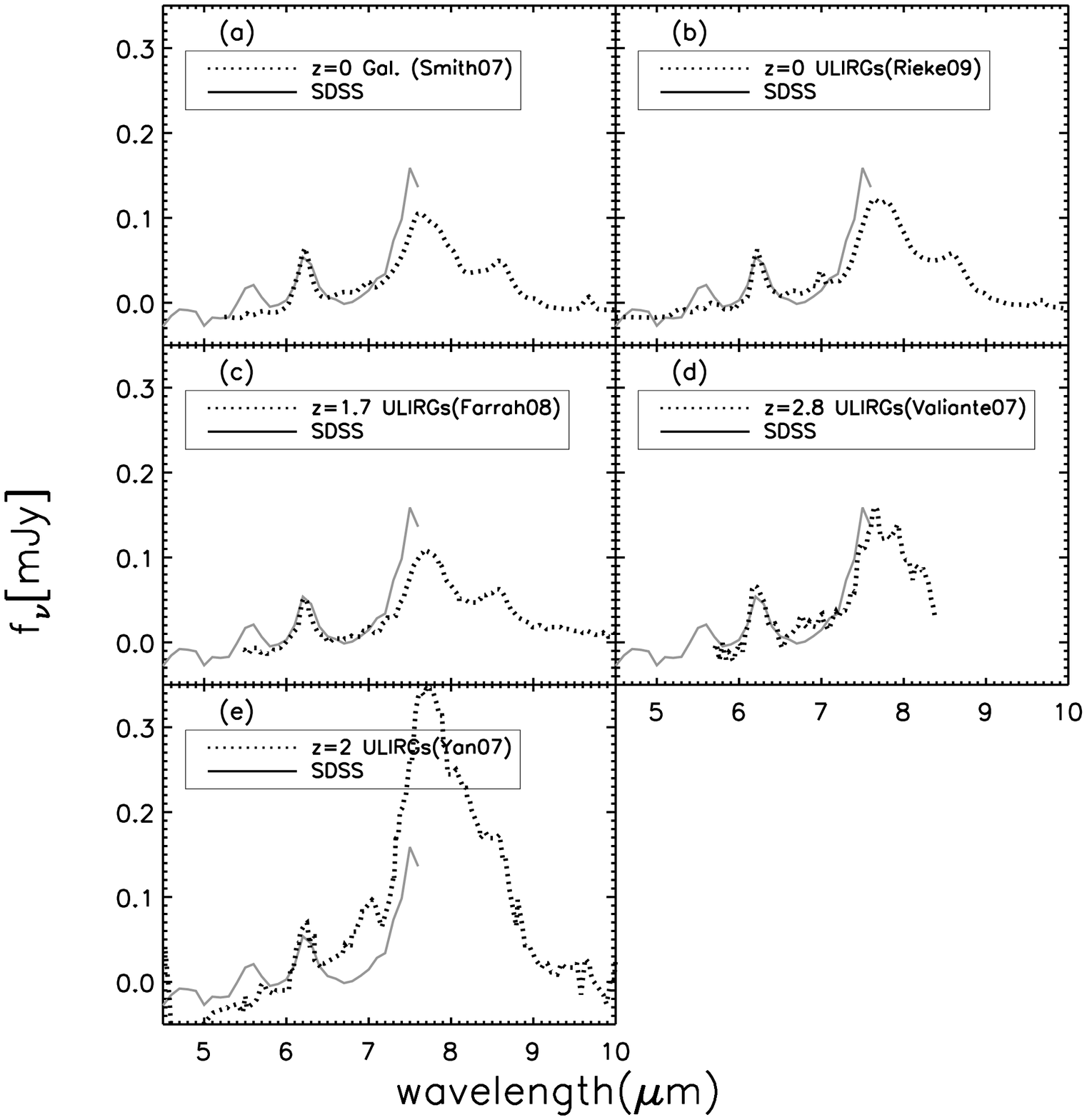}
\caption{\label{comp_SDSS_GAL}   The   continuum-subtracted  composite
spectrum of  the whole SDSS  sample (solid lines) compared  to those
(dotted  lines)  of  local  galaxies \citep{Smith07},  local  ULIRGs
\citep{Rieke09},  ULIRGs  at  $z=$1.7  \citep{Farrah08},  ULIRGs  at
$z$=2.8 \citep{Valiante07} and starburst-dominated ULIRGs at $z\sim$
2 from \citet{Yan07}.  }
\end{figure*}

\citet{Shi07} found  that the profiles  and relative strengths  of the
aromatic features  in local AGNs are  quite similar to  those of local
star-forming  galaxies,  implying  star  formation excitation  of  the
aromatic  features in  the AGNs.   Figure~\ref{comp_SDSS_GAL} compares
the continuum-subtracted  composite spectrum of the  whole SDSS sample
to   those   of   local   galaxies   \citep{Smith07},   local   ULIRGs
\citep{Rieke09}, ULIRGs at $z=$1.7 \citep{Farrah08}, ULIRGs at $z$=2.8
\citep{Valiante07}  and starburst-dominated ULIRGs  at $z\sim$  2 from
\citet{Yan07}.   The composite  spectra  in the  literature were  used
directly  while  the one  of  SDSS  quasars  was derived  through  the
arithmetic   mean  method  \citep[see][]{Shi07,   VandenBerk01}.   The
continuum of the SDSS composite  spectrum is determined by a power law
fit  to two  spectral regions  5.9-6.0  and 6.7-6.8  $\mu$m.  For  the
comparison, the spectra of other  samples were matched to the SDSS one
to   have  the   same  6.2   $\mu$m  aromatic   flux.   As   shown  in
Figure~\ref{comp_SDSS_GAL}(d), the ULIRGs of \citet{Valiante07} have a
$f$(7.7$\mu$mPAH)/$f$(6.2$\mu$mPAH) ratio consistent  with that of the
SDSS quasars.   A slightly higher  flux density around 7.7  $\mu$m for
the      SDSS     quasars      relative     to      local     galaxies
(Figure~\ref{comp_SDSS_GAL}(a)),              local             ULIRGs
(Figure~\ref{comp_SDSS_GAL}(b))  and  the  ULIRGs of  \citet{Farrah08}
(Figure~\ref{comp_SDSS_GAL}(c)) is most likely  due to the presence of
the    emission    line   [NeV]7.65$\mu$m    that    is   absent    in
star-formation-dominated  galaxies.   Only the  ULIRGs  at $z=2$  from
\citet{Yan07}   have   a   $f$(7.7$\mu$mPAH)/$f$(6.2$\mu$mPAH)   ratio
significantly  different  from the  SDSS  quasar  sample.  This  large
deviation is  at least partly  caused by absorption features  around 6
$\mu$m, such  as water ice  and hydrocarbons. Therefore,  the relative
strength of the 6.2 and  7.7 $\mu$m aromatic features most likely does
not change  as a function of  redshift or object  type.  This implies
star-formation excitation of the aromatic feature in the SDSS quasars.

\begin{figure*}
\epsscale{1.0}
\plotone{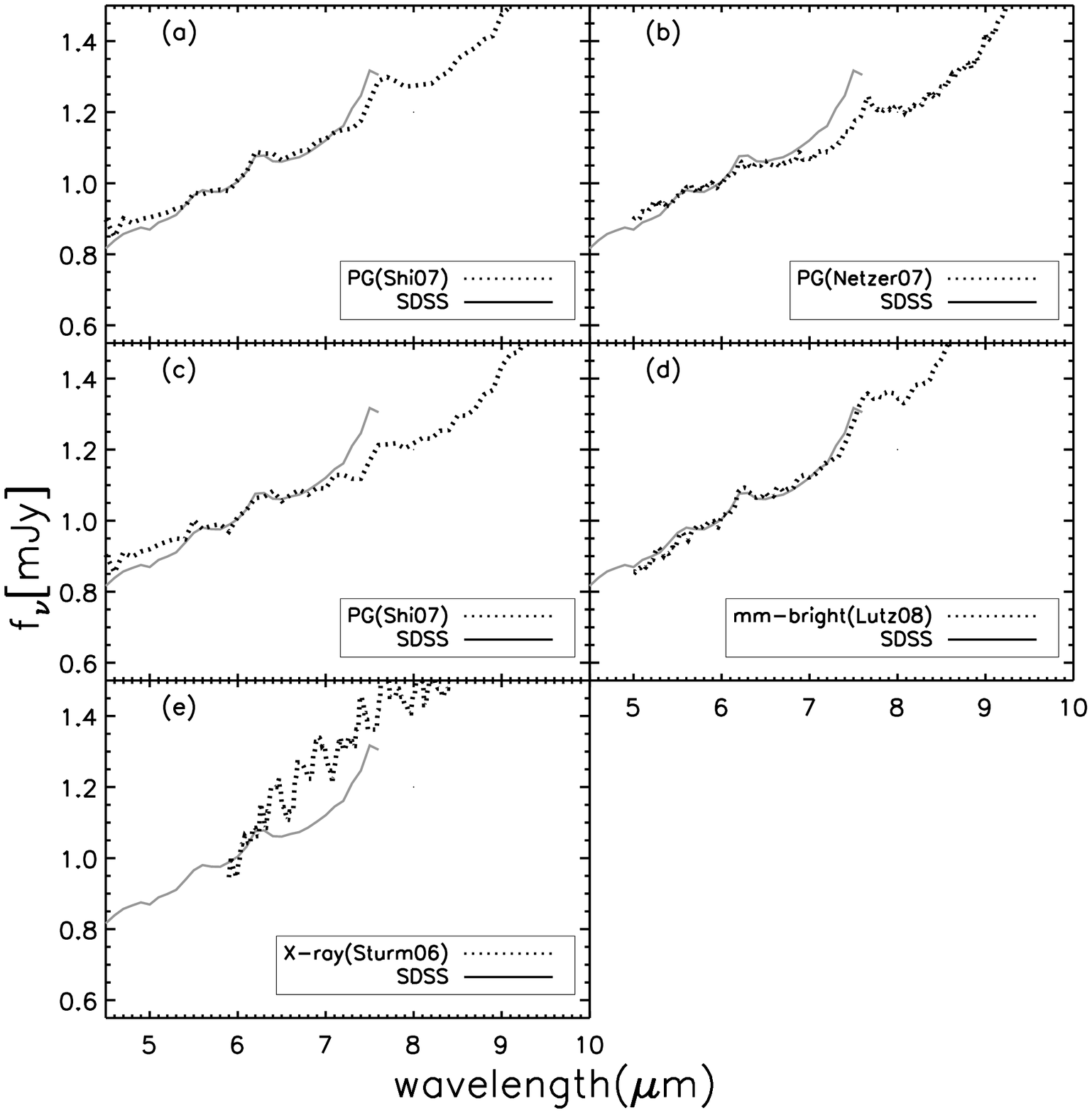}
\caption{\label{comp_SDSS_otherQSO} The composite spectrum (solid lines) of the whole SDSS sample
compared to those of other quasar samples (dotted lines).}
\end{figure*}
\
In Figure~\ref{comp_SDSS_otherQSO}, we  compare the composite spectrum
of the  whole SDSS sample  to those of  other quasar samples  that are
selected with different techniques and at different redshifts. For the
comparison,  all  spectra are  normalized  by  the  mean flux  density
between 5.8 and 6.1  $\mu$m.  The PG objects are UV/optically-selected
quasars  and thus resemble  the SDSS  quasars. The  comparison between
these two  samples should provide  clues to the redshift  evolution of
the     IR     properties     in    UV/optically-selected     quasars.
Figure~\ref{comp_SDSS_otherQSO}(a) compares  the composite spectrum of
the complete SDSS  sample to that of the complete  PG quasar sample at
$z$$\leq$0.5  from  \citet{Shi07}.   The  power-law slope  of  the  IR
continuum  is almost  the  same. The  EW  of the  6.2 $\mu$m  aromatic
feature is also similar in the  two samples.  In the spectral range of
the 7.7  $\mu$m feature,  the SDSS quasars  show slightly  higher flux
density,  which  may  be  caused  by stronger  silicate  emission,  IR
continuum   or   7.7   $\mu$m   feature   emission.    As   shown   in
Figure~\ref{comp_SDSS_otherQSO}(b), the  sub-sample of the  PG quasars
used in the QUEST project \citep{Netzer07, Schweitzer06} shows smaller
aromatic feature  EWs, which  is most likely  due to their  on average
higher nuclear  luminosity. To remove  the dependence of  the aromatic
feature EW on the quasar  luminosity, we compare the whole SDSS sample
to  the bright  PG sub-sample  that  is at  $M_{\rm B}$  $>$ -23  from
\citet{Shi07}.   The   median  5-6  $\mu$m   continuum  luminosity  is
7.5$\times$10$^{10}$ and  6$\times$10$^{10}$ L$_{\odot}$ for  the SDSS
and   the   bright  PG   sub-sample,   respectively.    As  shown   in
Figure~\ref{comp_SDSS_otherQSO}(c), the bright PG sub-sample has a 7.7
$\mu$m feature much weaker than  the SDSS quasars while its 6.2 $\mu$m
feature  is relatively  noisy but  most  likely weaker  than the  SDSS
quasars. This indicates that the fraction of the aromatic feature flux
in the  mid-IR evolves  with redshift at  a given  nuclear luminosity,
which  can be  caused by  either higher  SFRs in  SDSS quasars  or the
evolution of the aromatic feature  with redshift. As shown below, this
evolution is  actually dominated by the second  effect, i.e., stronger
aromatic flux for a given SFR at high redshift.

In Figure~\ref{comp_SDSS_otherQSO}(d), we  compare the SDSS quasars to
the mm-bright  type 1 quasars  at $z$$\sim$2 from  \citet{Lutz08} that
have mid-IR continuum luminosities two orders of magnitude higher than
for  the SDSS  quasars.  The  IR continuum  slope and  EWs of  the two
aromatic features are similar to those of the SDSS quasars. This again
suggests  that  the aromatic  feature  profile  does  not change  with
redshift, luminosity or object type.

The  composite spectrum  of type  2 QSOs  selected in  the  X-ray from
\citet{Sturm06} is  shown in Figure~\ref{comp_SDSS_otherQSO}(e).  This
sample  spans the  redshift range  from  0.2 to  1.38 and  has a  mean
redshift of  0.75.  The IR continuum  is redder than that  of the SDSS
type-1 quasars. Although no  aromatic bands are detected, the spectrum
is so noisy  that even higher EWs than for the  SDSS quasars are still
possible.   In   addition,   the   median  $L_{5-6{\mu}m}$   of   this
X-ray-selected  sample   is  about  2.5$\times$10$^{11}$  L$_{\odot}$,
several  times higher  than that  of the  SDSS sample.   Therefore, in
contrast to \citet{Sturm06}, we argue  that the S/N of the spectrum is
not high enough to rule  out the possibility that these X-ray-selected
type 2 QSOs have significant star-formation activity.

\subsection{Star Forming Infrared Luminosity Function of Quasar Hosts at $z$=1}

\subsubsection{Conversion factors from the 6.2 $\mu$m aromatic flux to the total star-forming IR luminosity}

\begin{deluxetable*}{llllcllllll}
\tablecolumns{10}
\tabletypesize{\scriptsize}
\tablewidth{0pc}
\tablecaption{\label{HIG-Z-SB} High-Redshift Star-Forming Galaxies}
\tablehead{
  \colhead{Sources}        & \colhead{z}                 &  \colhead{$F_{16{\mu}m}$}     & 
  \colhead{$F_{24{\mu}m}$}  & \colhead{$F_{70{\mu}m}$}     &  \colhead{$F_{160{\mu}m}$}    &
  \colhead{Ref.}           &  \colhead{$F_{6.2{\rm PAH}}$} & \colhead{EW$_{6.2{\rm PAH}}$} &
  \colhead{$L_{\rm TIR}$}   \\
  \colhead{   }            & \colhead{   }              &   \colhead{[mJy]}            & 
  \colhead{[mJy]}          &  \colhead{[mJy]}           &   \colhead{[mJy]}           &
  \colhead{ }              & \colhead{[$10^{-14}$erg/s/cm$^{2}$]}               &   
  \colhead{[${\mu}m$]}     & \colhead{[$10^{11}$L$_{\odot}$]}  \\
  \colhead{(1)}            & \colhead{(2)}            &  \colhead{(3)}                 &   
  \colhead{(4)}            & \colhead{(5)}            &  \colhead{(6)}                 &
  \colhead{(7)}            & \colhead{(8)}            &  \colhead{(9)}                 &
  \colhead{(10)} }
\startdata
Brand08-70bootes3  &  0.986  & --              &  1.26$\pm$0.1     & 35.0$\pm$6.3  & 145$\pm$29  & 1 & 0.96$\pm$0.07  &  0.48 & 46.39\\
Brand08-70bootes4  &  0.975  & --              &  1.22$\pm$0.1     & 36.6$\pm$5.1  & 100$\pm$20  & 1 & 0.84$\pm$0.08  &  0.40 & 41.29\\
Brand08-70bootes7  &  0.664  & --              &  2.18$\pm$0.1     & 51.8$\pm$4.7  & 135$\pm$27  & 1 & 1.66$\pm$0.08  &  0.51 & 20.20\\
Brand08-70bootes9  &  0.668  & --              &  3.46$\pm$0.1     & 67.2$\pm$3.1  & 245$\pm$49  & 1 & 3.29$\pm$0.12  &  0.79 & 29.68\\ 
Murphy09-ID3       &  0.63   & 0.774$\pm$0.006 &  1.210$\pm$0.005  & 11.1$\pm$0.53 & --          & 2 & 1.18$\pm$0.03  &  0.72 &  6.60\\    
Murphy09-ID8       &  0.64   & 0.399$\pm$0.006 &  0.721$\pm$0.005  & 11.1$\pm$0.53 & --          & 2 & 0.56$\pm$0.02  &  0.49 &  4.93\\    
Murphy09-ID11      &  1.22   & 0.993$\pm$0.006 &  0.446$\pm$0.005  & 13.2$\pm$0.53 & --          & 2 & 0.76$\pm$0.03  &  0.73 & 53.28\\    
Murphy09-ID22      &  0.64   & 0.580$\pm$0.006 &  0.750$\pm$0.005  & 5.53$\pm$0.53 & --          & 2 & 0.89$\pm$0.03  &  0.58 &  4.16\\    
Teplitz07-2        &  1.09   & --              &  0.133$\pm$0.015  & --            & --          & 3 & 0.16$\pm$0.03  &  0.70 &  3.82\\

\enddata
\tablecomments{
References: 1 -- \citet{Brand08}; 2 -- \citet{Murphy09}; 3 -- \citet{Teplitz07} }
\end{deluxetable*}

\begin{figure}
\epsscale{1.2}
\plotone{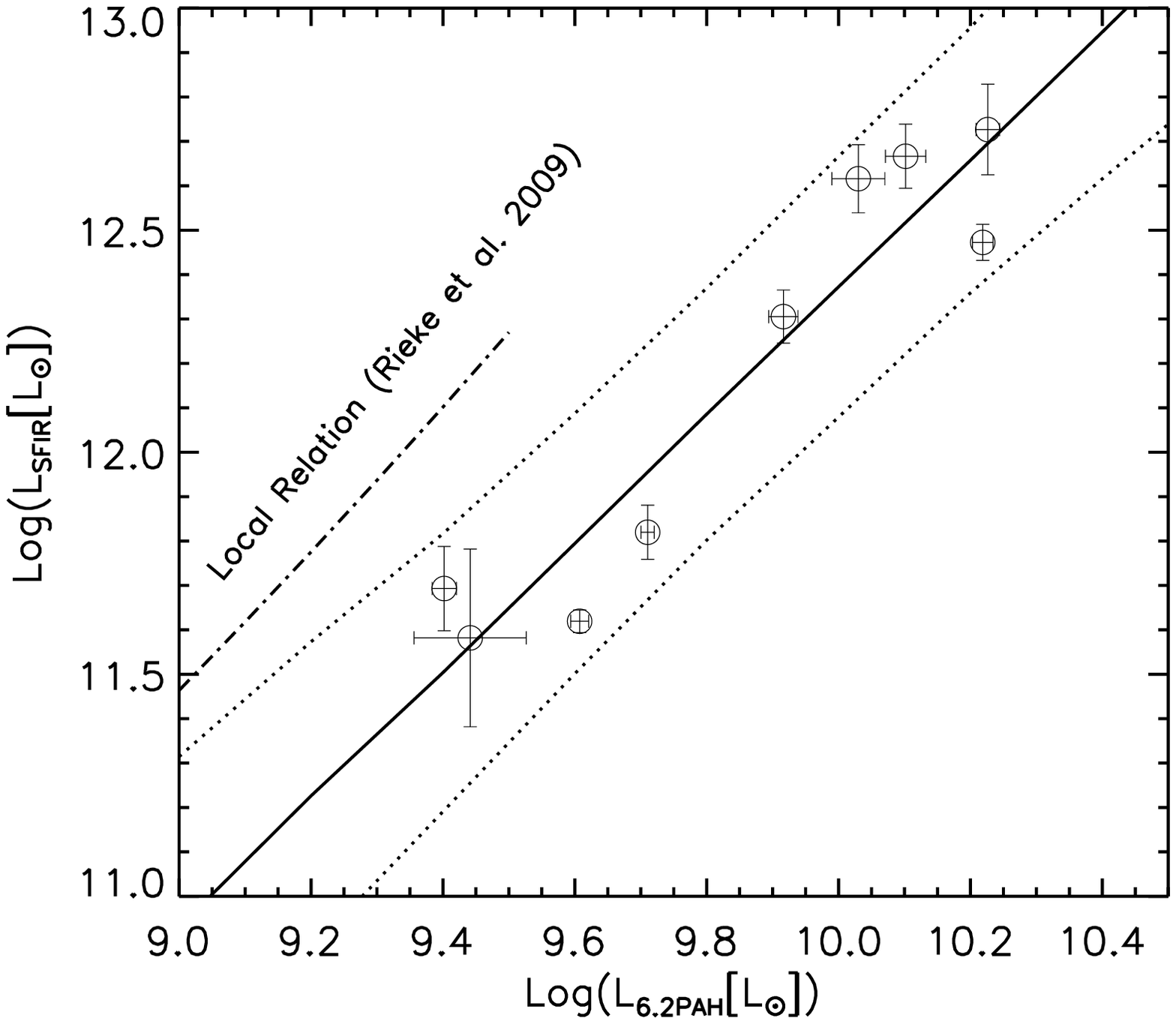}
\caption{\label{PAH_TIR_SB} The $L_{\rm   SFIR}$
vs. $L_{\rm 6.2{\mu}mPAH}$ for high-redshift star-forming galaxies as listed
in Table~\ref{HIG-Z-SB}. The solid line and dotted line show the linear regression fit to the data and 
the  associated scatter, respectively. The dot-dashed line shows the local relationship based on
the star-formation template of \citet{Rieke09}.}
\end{figure}

High-redshift star-forming  galaxies appear  to show evolution  in the
relative  strengths of  the  total star-forming  IR  and the  aromatic
luminosities  ($L_{\rm SFIR}/L_{\rm PAH}$)  \citep{Rigby08, Murphy09}.
To convert the aromatic flux of quasar hosts to the corresponding SFR,
the ratio ($L_{\rm SFIR}/L_{\rm PAH}$)  should be defined based on
high-redshift  star-forming  galaxies.    A  heterogeneous  sample  of
high-redshift star-forming galaxies at 0.5 $<z<$ 1.5 was compiled from
the  literature, as  listed  in Table~\ref{HIG-Z-SB}.   The object  is
named  by the  corresponding reference  followed  by the  name in  the
reference.   The  IRS spectra  were  retrieved  from  the archive  and
re-reduced except for Teplitz07-2 for which the published spectrum was
used.  Since  the aromatic fluxes measured with  different methods can
vary up to  several times \citep{Smith07}, the 6.2  $\mu$m features of
these star-forming  galaxies were measured through the  same method as
for the quasars.  The large  equivalent width ($\gtrsim$ 0.3 $\mu$m) of
the  6.2  $\mu$m  feature  in  these objects  indicates  the  dominant
component of  the IR radiation is  powered by star  formation. This is
derived  from  the  fact  that  the  composite  HII-like  star-forming
galaxy spectra in \citet{Smith07} have EW(6.2$\mu$mPAH) of $\sim$0.3 $\mu$m.

To  derive the total  IR luminosity  of these  star-forming galaxies,
their broad-band IR flux densities are fitted with the 14 star-forming
templates  for the  IR  luminosity range  of  $10^{9.75}$ -  $10^{13}$
L$_{\odot}$ from  \citet{Rieke09}.  As shown  in Table~\ref{HIG-Z-SB},
the  MIPS  24,  70  and   160  $\mu$m  photometry  are  used  for  the
\citet{Brand08} sample, while IRS 16 $\mu$m, MIPS 24 and 70 $\mu$m are
used for the \citet{Murphy09} sample. The $L_{\rm SFIR}$(8-1000$\mu$m)
is  measured   by  integrating   the  template  giving   the  smallest
$\chi$$^{2}$. For each object,  the observed photometry and associated
uncertainties  are  Monte-Carlo   simulated  and  re-fitted  with  the
star-forming  templates.   The  standard  deviation of  the  resulting
$L_{\rm SFIR}$(8-1000$\mu$m) is adopted as the 1-$\sigma$ error of $L_{\rm
  SFIR}$ due to the  observed flux uncertainties.  An additional error in
$L_{\rm SFIR}$ due to the  scatter of the template itself is estimated
as ($L_{\rm SFIR}^{\rm template  i+1}$ - $L_{\rm SFIR}^{\rm template i-1}$)/2,
where the template $i$ is the one giving the minimum $\chi$$^{2}$. The two
errors are added quadratically to  give the final error in $L_{\rm
  SFIR}$.  \citet{Murphy09}  have shown that the  $L_{\rm SFIR}$ using
photometry  including MIPS  70  or  160 $\mu$m  flux  densities is  an
unbiased estimate of the total  IR luminosity based on photometry from
the near-IR to the sub-mm.   For one object (Teplitz07-2) with only 24
$\mu$m data, its  24 $\mu$m flux density is converted  to the total IR
luminosity using  the redshift-dependent $L_{24{\mu}m}$-$L_{\rm SFIR}$
relationship  and the  1-$\sigma$ scatter  is  assumed to  be 0.2  dex
\citep{Rieke09}.  \citet{Murphy09} show that,  at $z\sim$ 1, the total
IR  luminosity estimated  from the MIPS  24 $\mu$m  flux density  does not
suffer from  any systematic  shift because the aromatic
flux contributes a small fraction at rest-frame 12 $\mu$m.

Figure~\ref{PAH_TIR_SB} shows the  $L_{\rm SFIR}$ vs. $L_{\rm PAH}$ for
these high-redshift star-forming galaxies.  As a comparison, the local
relationship  is   shown  as  the   dash-dotted  line  based   on  the
star-forming template  of \citet{Rieke09}.  There  is a factor  of 2-3
decrease in the $L_{\rm  SFIR}$/$L_{\rm 6.2PAH}$ at $z\sim$ 1 compared
to  the  local  behavior.  We  used  the  IDL  code  linmix$\_$err.pro
\citep{Kelly07}  that  employs  a Bayesian  approach to  perform  linear
regression of two variables with measured errors. The following relationship is derived
for quasar hosts at $z$ $\sim$ 1:
\begin{eqnarray}
\mbox{Log}(L_{\rm SFIR}) & = & (12.16\pm0.09)+ \nonumber \\
& &(1.42\pm0.29)({\rm Log}(L_{\rm 6.2PAH}) - 9.85) \nonumber \\
& & \pm0.25
\end{eqnarray}
Note  that  the relationship  is  obtained  between ${\rm  Log}(L_{\rm
  SFIR})$ and $L_{\rm 6.2PAH} - {\langle}L_{\rm 6.2PAH}{\rangle}$.  In
this case, the slope and  intercept of the relationship are independent
of each other. The median  and 1-$\sigma$ range of the regression line
are   shown   as   solid    and   dotted   lines,   respectively,   in
Figure~\ref{PAH_TIR_SB}.  As shown  in Table~\ref{HIG-Z-SB}, we do not
have  star-forming galaxies at  $z\sim$ 1  with total  IR luminosities
below 10$^{11}$  $L_{\odot}$.  It thus  may be invalid  to extrapolate
this  relation  below  10$^{11}$  $L_{\odot}$.  

As  our  objects have  LIRG-level  SFIR  luminosities (Table~\ref{PAH_Table}), the  conversion
factor $L_{\rm SFIR}/L_{\rm PAH}$  is about three times lower compared
to  the local relationship.  Therefore, our  estimate of  evolution of
star formation in quasar hosts  is conservative. If we adopt the local
relationship  for our  $z$ $\sim$  1  quasars, as  some works  provide
tentative  evidence for  no  evolution in  $L_{\rm SFIR}/L_{\rm  PAH}$
ratio \citep[e.g.][]{Magnelli09}, then the evolution from $z$ $\sim$ 1
to $z$  $\sim$ 0 would be increased  by a factor of  about three above
our estimate.

\subsubsection{Star-forming IR luminosity function of $z\sim$ 1 Type-1 Quasars}\label{SFLF-QSO}

\begin{figure*}
\epsscale{1.0}
\plotone{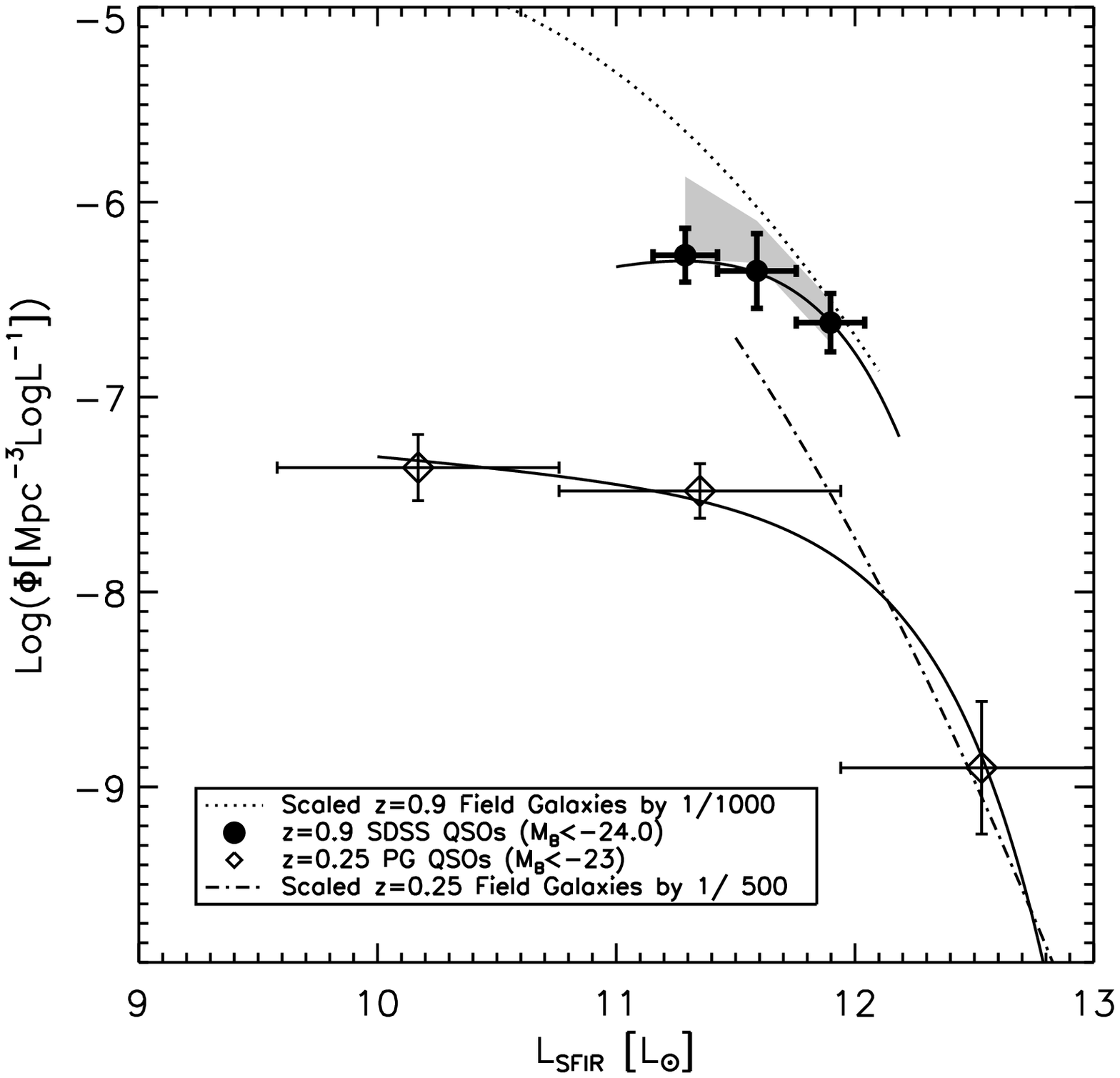}
\caption{\label{Lum_Func}  The  star-forming  IR  luminosity  function
(SFIR LF)  of $z\sim$  1 quasar hosts  (filled circles)  compared to
that of low-redshift PG quasars (open diamonds). The grey area shows
the 1-$\sigma$ range of the  SFIR LF after correcting the effect due
to the uncertainty of the SFIR luminosity (see text for details).
The SFIR  LFs of field galaxies  at both low and  high redshifts are
also shown for comparison.}
\end{figure*}

The SFIR LF  of the $z\sim$ 1 type-1 quasar hosts  is constructed by counting
the number of  objects in different luminosity bins  and redshift bins
(${\Delta}L{\Delta}z$).  For this method, in the lowest luminosity bin
that intersects the flux limit at a given redshift, only the part
of the ${\Delta}L{\Delta}z$  space above the flux limit  can be filled
by objects.   To account for  this partly filled  bin, we have  used a
revised 1/$V_{\rm max}$ method \citep{Page00} to derive the luminosity
function.   The incompleteness  function for  the quasar  selection is
given by \citet{Richards06}. The luminosity bins are defined to have roughly
the same number of objects and range from the observed lowest luminosity to the
highest one.

Figure~\ref{Lum_Func}  shows the SFIR  LF of  $z\sim$ 1  type-1 quasar
hosts  compared  to  the   LF  of  contemporary  field  galaxies  from
\citet{LeFloch05} and to that of  the $z$=0.25 PG quasars. The SFIR LF
of the PG  quasars has been updated by  using the $L_{\rm SFIR}/L_{\rm
  PAH}$ ratio  of \citet{Rieke09}, but  there is only a  small change.
The error bar  shows the Poisson uncertainty.  The  uncertainty in the
SFIR luminosity  can smooth the LF  by scattering more  objects out of
the  luminosity bin  originally with  a larger  number of  objects. To
quantify  this  effect,  a  set  of 1000  SFIR  LFs  were  Monte-Carlo
simulated   using   the  observed   $L_{\rm   SFIR}$  and   associated
uncertainties.  The relative  change in  the number  density  for each
luminosity  bin is  then  quantified.  The grey  area  then shows  the
1-$\sigma$ range of  the SFIR LF after correcting  for the effects due
to the uncertainty of the SFIR luminosity.

\citet{Shi07} carried out a  series of simulations to demonstrate that
the PG quasars have a flat SFIR LF compared to that of field galaxies.
As shown in the figure, such  a trend most likely exists at $z\sim$ 1.
Within our limited range of luminosity, the difference between the $z$
$\sim$ 0.25 and $z$ $\sim$ 1 SFIR LFs can be described as pure density
evolution.

\subsubsection{Cosmic Evolution of the Comoving SFIR Luminosity Density And Average SFR  in Quasar Hosts}\label{EVL-SFR}

\begin{deluxetable*}{ccccccccccccccccc}
\tabletypesize{\scriptsize}
\tablecolumns{7}
\tablecaption{\label{Best_Fit} Best-fitting parameters to star-forming IR LF of SDSS and PG quasars}
\tablewidth{0pt}
\tablehead{ \colhead{Object}                        & \colhead{Log($\phi^{\star}$[Mpc$^{-3}$ LogL$^{-1}$]) } & 
            \colhead{Log($L^{\star}$[L$_{\odot}$])} & \colhead{$\alpha$}                                       }
\startdata
SDSS            & -6.30$\pm$0.16 &     11.64$\pm$ 0.19 &  -0.59$\pm$0.25 \\
PG($M_{B}<$-23) & -7.86$\pm$0.39 &     12.04$\pm$0.34  &  -1.10$\pm$0.19 \\
\enddata
\tablecomments{
The slope for the SDSS sample is derived by producing the observed PAH
detection   rate   (see  Figure.~\ref{DecRate_AsFuncOfSlope}), while the
one for the PG sample is calculated using the number  densities  and associate
uncertainties in two low luminosity  bins. By Monte-Carlo  simulating the  derived slope,  the  $\phi^{\star}$ and
$L^{\star}$  is  obtained  by  fitting Schechter  functions  to  the
observed LFs. All uncertainties are given at 1-$\sigma$ level.}
\end{deluxetable*}

\begin{figure}
\epsscale{1.2} 
\plotone{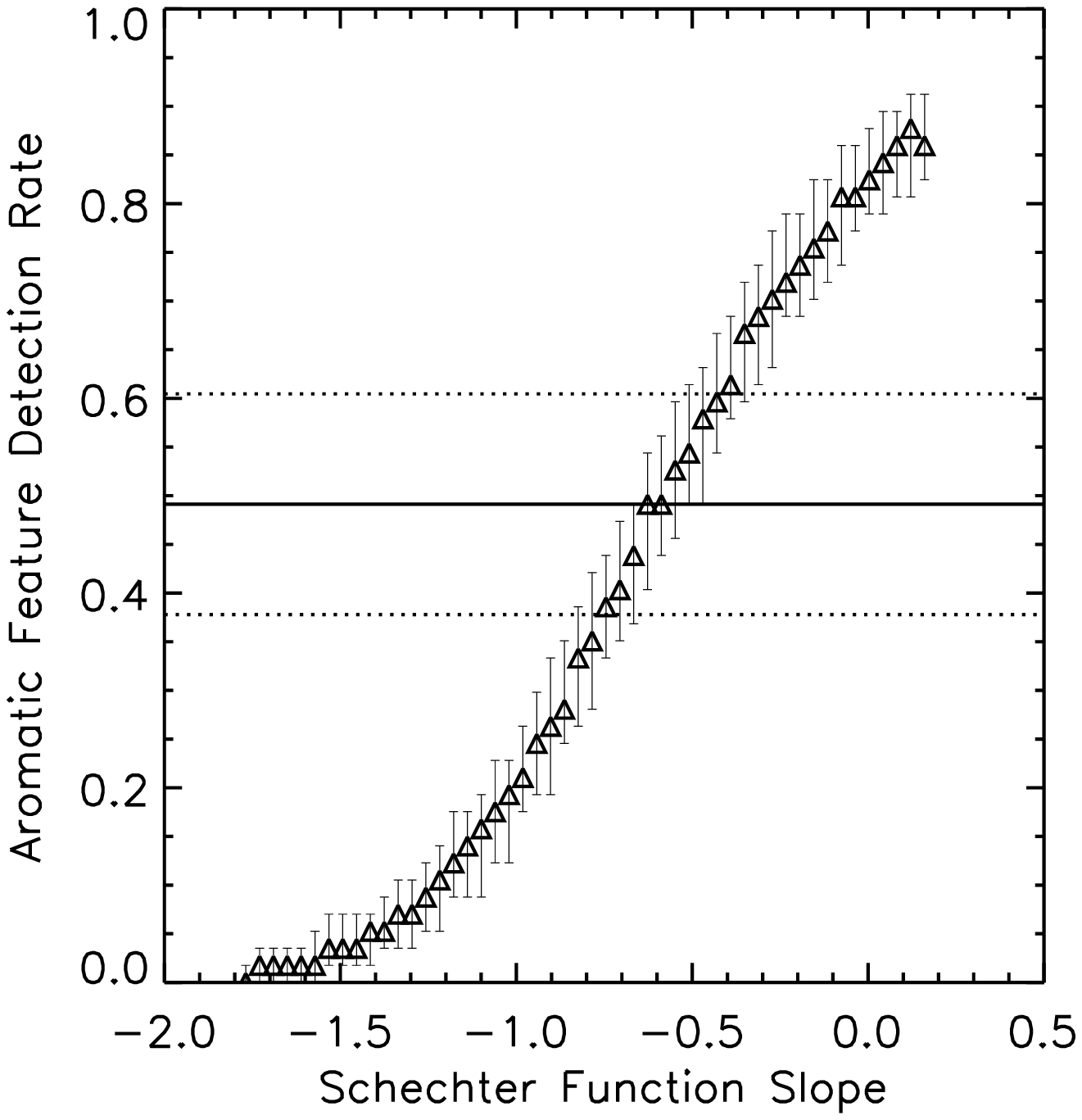}
\caption{ \label{DecRate_AsFuncOfSlope} The triangle shows the 
predicted detection rate of the aromatic feature of the SDSS sample 
for a range of the assumed Schechter function slope. The solid line is
the observed detection rate while the associated uncertainty is shown as
the dotted line.}
\end{figure}

\begin{figure}
\epsscale{1.3}
\plotone{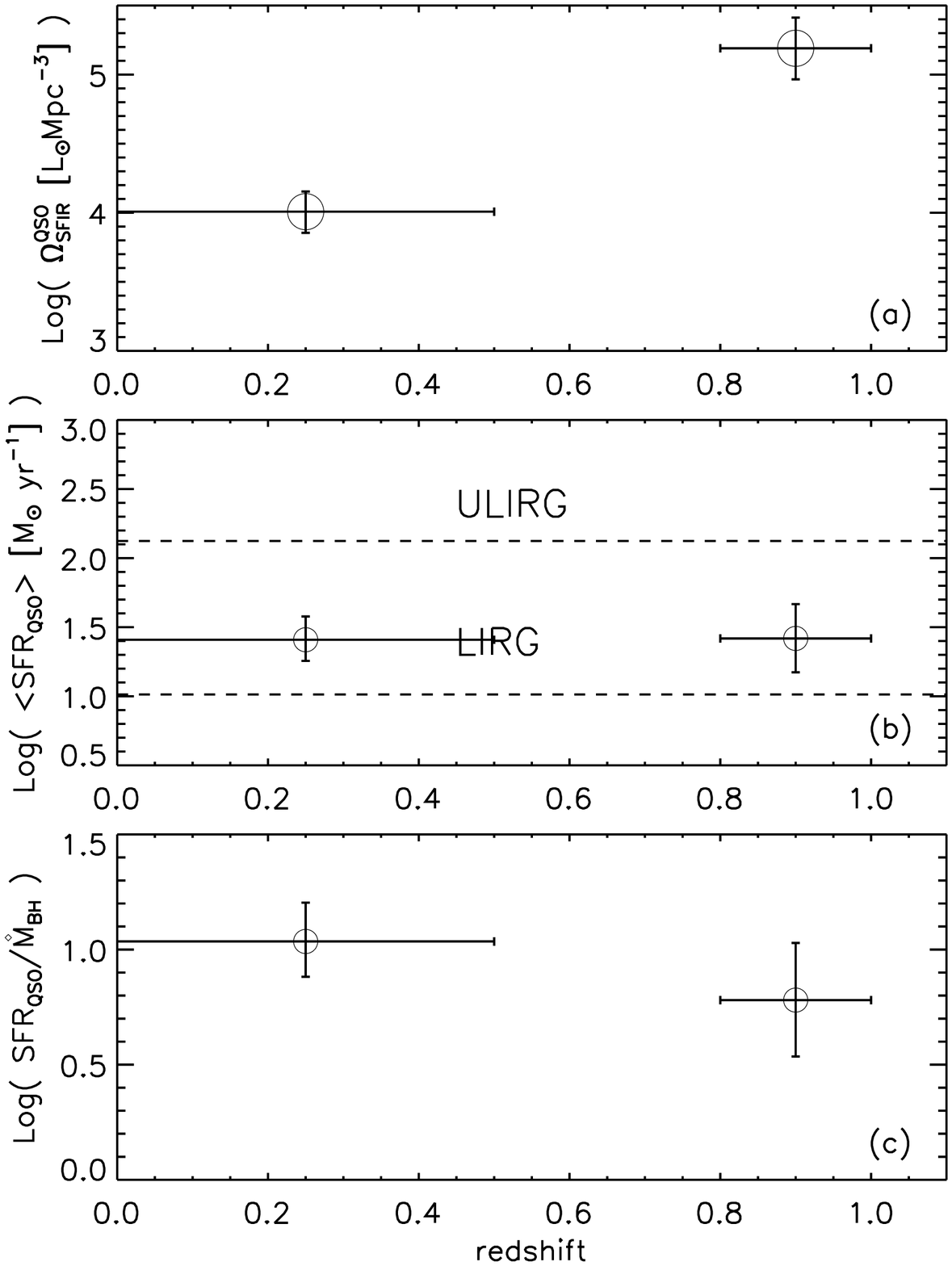}
\caption{ \label{EVL_SF_QSO}  The cosmic evolution of the comoving SFIR luminosity
density (a), the average SFR (b) and the average SFR/$\dot{M}_{\rm acc}$ (c) in type 1 quasar hosts.
The SFIR luminosity is converted to the SFR using \citet{Rieke09} relations that deviate from
those of \citet{Kennicutt98} by a factor of $\sim$2 as a result of a different initial mass function. }
\end{figure}

With the  SFIR LF  at $z$  $\sim$ 0.25 and  $z$ $\sim$  1, we  now can
quantify the cosmic evolution  of the comoving SFIR luminosity density
in type-1 quasar hosts. Due to the lack of constraints on the slope at
low-luminosity, two types of estimation have been explored.  First, we
can  estimate the  comoving  SFIR density  by  simply integrating  the
available data  points, which gives  an evolution factor  of 10$\pm$4.
Second, we  can fit the SFIR  LF with a series  of Schechter functions
whose slopes ($\alpha$) are Monte-Carlo simulated.  For the PG sample,
the  slopes are  simulated using  the number  densities  and associate
uncertainties in two low luminosity  bins. For the SDSS sample, as the
slopes  given in  a  similar  way have  large  uncertainties, we 
attempt  to  use the  objects  with  undetected  aromatic features  to
constrain  the slope  $\alpha$.   Basically, for  a  range of  assumed
slopes, we fitted data points to obtain a series of analytic Schechter
LFs. For  each LF, a  SFIR luminosity is  then randomly assigned  to a
SDSS quasar  with the relative  probability following that of  the LF.
The detection rate of the aromatic feature is then calculated by comparing
the simulated  SFIR luminosity to the  observed 3-$\sigma$ uncertainty
of the SFIR  luminosity.  Figure~\ref{DecRate_AsFuncOfSlope} shows the
simulated aromatic  detection rates (points) compared  to the observed
one         (lines).         The        derived         slope        is
$\alpha=-0.59{\pm}0.25$.  By  Monte-Carlo  simulating  this
slope, the observed LF is  fitted with a series of Schechter functions,
which is shown as a  solid line in Figure~\ref{Lum_Func} and listed in
Table~\ref{Best_Fit}.    We   obtained   Log$L^{*}$   =
11.64$\pm$0.19.  The comoving SFIR luminosity density is then obtained
by integrating the  fitted Schechter functions and the  result is shown
in  the  Figure~\ref{EVL_SF_QSO}(a).   In  this  method,  the  derived
increase of the comoving SFIR  luminosity density in quasar hosts from
z=0.25 to  z=0.9 is 15$^{+7}_{-5}$  (1-$\sigma$).  In either case,  the SFIR
energy density in type 1 quasar hosts shows a dramatic evolution, much
larger than the behavior of the general field galaxies \citep[which increase
by   a  factor   of  5$^{+1}_{-0.7}$; ][]{LeFloch05,   Perez-Gonzalez05}. 

The integration  of  the  LF  is  carried  out  in  the  luminosity  range
[10$^{10}$,  10$^{12}$]   L$_{\odot}$.   If we  vary   the  upper-  or
lower-limits by a  factor of 10, the derived  evolution changes little.
If we had used the local  conversion of the aromatic luminosity to the
SFR, the evolution of the quasar host SFRs would roughly triple.

We now consider  how selection biases might influence  this result. We
quantified the  evolution of the  comoving SFIR luminosity  density in
quasar hosts by measuring the SFR  in PG quasars and SDSS quasars, two
samples  selected  through different  UV/optical  criteria.  The  SDSS
quasars were  selected on the  basis of multiple  color-color diagrams
\citep{Richards01}; nonetheless,  as with the PG sample,  UV excess is
the  dominant  characteristic. The  incompleteness  of  the PG  quasar
selection  (U-B  $<$  -0.46 and  B  $<$  16.16)  has long  been  known
\citep{Goldschmidt92,     Wisotzki00,     Mickaelian01,     Jester05}.
Comparisons  of  PG  quasars   to  other  quasar  surveys  indicate  a
completeness   of   $\sim$50-100\%  \citep{Wisotzki00,   Mickaelian01,
  Jester05}.  The incompleteness is  most likely independent  of the
quasar  optical and radio  properties within  the PG  quasar selection
criterion itself \citep{Jester05}, i.e., PG quasars are representative
of quasars that are bright at B-band.  Therefore, the evolution of the
comoving  SFIR   luminosity  density  in   type  1  quasar   hosts  is
over-estimated at most  by a factor of 2.  In  addition, our $z\sim$ 1
quasars ($M_{i}<$  -24.7; $M_{B}<\sim$ -  24.0) are brighter  than our
low-redshift PG quasars by one magnitude ($M_{B}<$ -23).  Our study of
PG quasars  showed that  brighter PG quasars  have a flatter  SFIR LF,
implying  on average  a stronger  SFR in  a brighter  PG  quasar host.
However, the integral SFIR density is  actually lower by a factor of 2
for PG quasars  at $M_{B}<$ -23 compared to  the value for $M_{B}<$ -22,
simply due to  lower number densities.  The evolution  of the comoving
SFIR density in quasar hosts may  be even larger if we compare quasars
at the same brightness cut. Therefore, the two factors (in-completeness of
the PG sample  and un-matched brightness limit of  two quasar samples)
 tend to cancel each other out. We conclude that the  evolution of  the  comoving SFIR
luminosity density in type-1 quasar  hosts is indeed much larger than that in
field galaxies.

We can use the SFIR LF of quasar hosts to
derive the average SFR per object:
\begin{equation}
{\langle}{\rm SFR}_{\rm qso}{\rangle}=\frac{\int_{0}^{\infty} {\rm SFR}_{L_{\rm SFIR}}\Phi_{\rm SFIR}dL_{\rm SFIR}}
{\int_{L_{\rm qso}^{\rm limit}}^{\infty} \Phi_{\rm qso}dL_{\rm qso}},
\end{equation}
where $\Phi_{\rm SFIR}$ is the  SFIR LF of quasar hosts and $\Phi_{\rm
  qso}$ is the LF of the quasar luminosity at the wavelength where the
quasar    sample   is    selected.    The    result   is    shown   in
Figure~\ref{EVL_SF_QSO}(b), where  $L_{\rm SFIR}$ is  converted to the
SFR                 using                 the                 relation
SFR($M_{\odot}$/yr)=6.37$\times$10$^{-12}$$(L_{\rm
  TIR}/L_{\odot})^{1.11}$ at  $L_{\rm TIR}$ $>$  10$^{11}$ L$_{\odot}$
and                  SFR($M_{\odot}$/yr)=2.31$\times$10$^{-11}$$(L_{\rm
  TIR}/L_{\odot})^{1.06}$   at    $L_{\rm   TIR}$   $\leq$   10$^{11}$
L$_{\odot}$ \citep{Rieke09}. These relationships deviate from those of
\citet{Kennicutt98} mainly  due to a different  initial mass function.
As shown in Figure~\ref{EVL_SF_QSO}(b),  the average SFR in the quasar
host  is LIRG-level  (and remains  in the  LIRG range  with  the local
$L_{\rm SFIR}/L_{\rm PAH}$ applied  at $z$ $\sim$1).  The evolution of
this average  SFR is very  insignificant, and remains small  even with
application of the  local $L_{\rm SFIR}/L_{\rm PAH}$ at  $z$ $\sim$ 1.
As we  noted above, our SDSS  quasar sample is limited  at a magnitude
($M_{B}<$-24)  brighter  than   the  PG  quasar  limit  ($M_{B}<$-23).
Brighter  quasars  have  on  average higher  SFRs  \citep{Shi07}.   PG
quasars at  $M_{B}<$ -23 have SFRs  on average a factor  of 1.3 higher
than PG  quasars at  $M_{B}<$ -22. Therefore,  the average SFR  in the
optical type-1  quasar is most  likely nearly constant  with redshift.
This result supports our conclusion  that most of the host galaxy SFIR
LF evolution is in density,  not in luminosity.  Here we simply assume
that the same SFR-$M_{\rm B}$ trend  holds at z=0 and z=1 and that the
trend  between $M_{\rm  B}$=-22 to  -23 holds  up to  $M_{\rm B}$=-24,
which is  most likely  true based on  the correlation between  SFR and
quasar nuclear luminosity  as discussed in \S~\ref{SFIR-NUCLUM}.  Note
that the  derived quantities in Figure~\ref{EVL_SF_QSO}  do not depend
on the integration  limit.  For example, the change  is only about 0.1
dex  using   the  observed  range  of  the   SDSS  sample  ($[10^{11},
  10^{12}]$).

\subsection{The SFIR Luminosity As A Function of Quasar Nuclear Luminosity}\label{SFIR-NUCLUM}

\begin{deluxetable*}{ccccccccccccccccc}
\tabletypesize{\scriptsize}
\tablecolumns{7}
\tablecaption{\label{Fit_LNUC_LSFIR} Best-fitting parameters to the correlation $L_{\rm SFIR}$-$L^{\rm NUC}_{5-6{\mu}m}$ }
\tablewidth{0pt}
\tablehead{ \colhead{data sets}    & \colhead{Best-fitting Formula} & \colhead{correlation coefficient}  }
\startdata
All PG + SDSS                    & ${\rm Log}(L_{\rm SFIR}) = (10.98\pm0.05)+ (0.97\pm0.08)({\rm Log}(L_{\rm 5-6{\mu}m}) - 10.58)\pm0.52$  &  0.76$\pm$0.05\\
All detected plus stacked points & ${\rm Log}(L_{\rm SFIR}) = (11.00\pm0.07)+ (1.14\pm0.11)({\rm Log}(L_{\rm 5-6{\mu}m}) - 10.42)\pm0.56$  &  0.81$\pm$0.05\\
All PG                           & ${\rm Log}(L_{\rm SFIR}) = (10.54\pm0.12)+ (0.86\pm0.15)({\rm Log}(L_{\rm 5-6{\mu}m}) - 10.37)\pm0.75$  &  0.64$\pm$0.11\\
\enddata
\tablecomments{The correlations are derived between Log($L_{\rm SFIR}$) and 
Log$L^{\rm NUC}_{5-6{\mu}m}$ - mean($L^{\rm NUC}_{5-6{\mu}m}$). In this case,
the uncertainties of the intercept and slope are independent.}
\end{deluxetable*}

\begin{figure*}
\epsscale{1.}
\plotone{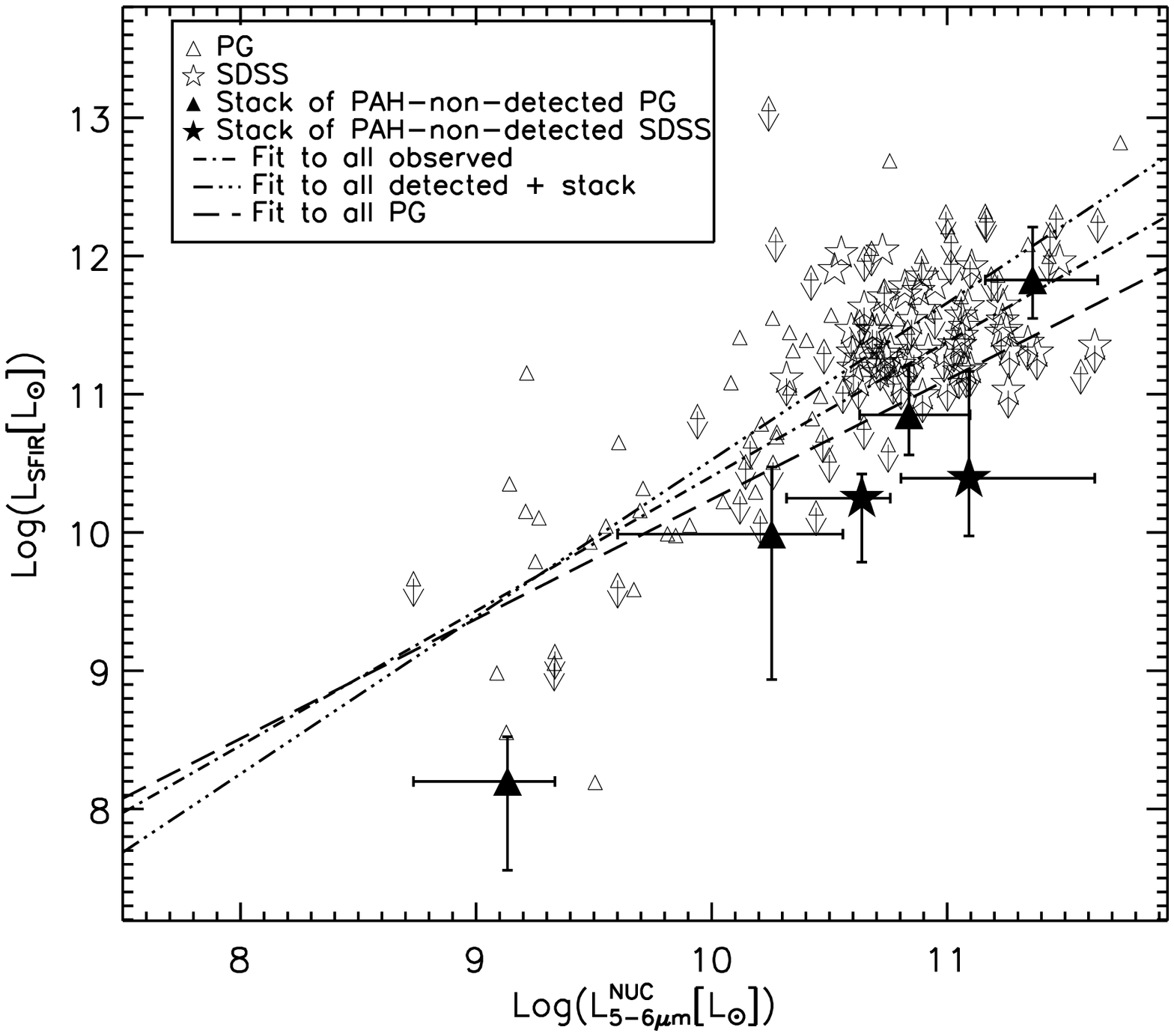}
\caption{ \label{LSFIR_LQSO} The correlation between the star-forming IR luminosity and
the nuclear luminosity at 5-6 $\mu$m. The open triangles and stars are for individual
PG and SDSS quasars, respectively.  The large filled symbols  show the result based 
on the composite spectra of PAH-undetected objects within 5-6 $\mu$m luminosity 
ranges as indicated by their error-bars on the X-axis (see Figure~\ref{COMP_SPEC}). Three linear regression fits are 
listed in  Table~\ref{Fit_LNUC_LSFIR}.}
\end{figure*}

The SFIR  LF of  the type-1 quasar  hosts gives  a global sense  of their
star-forming  activity. The  enhanced SFR  in the
quasar  hosts as  implied by  the SFIR  LF indicates  that  further
investigation of the star  formation activity in individual objects is
critical  to  understand  the  interplay between  star  formation  and
nuclear activity.  As shown  in \citet{Shi07}, the brighter PG quasars
have on  average higher SFRs as  measured from their SFIR  LF.  We now
search for direct correlations  between the SFIR luminosity and quasar
nuclear luminosity.

\begin{figure}
\epsscale{1.2}
\plotone{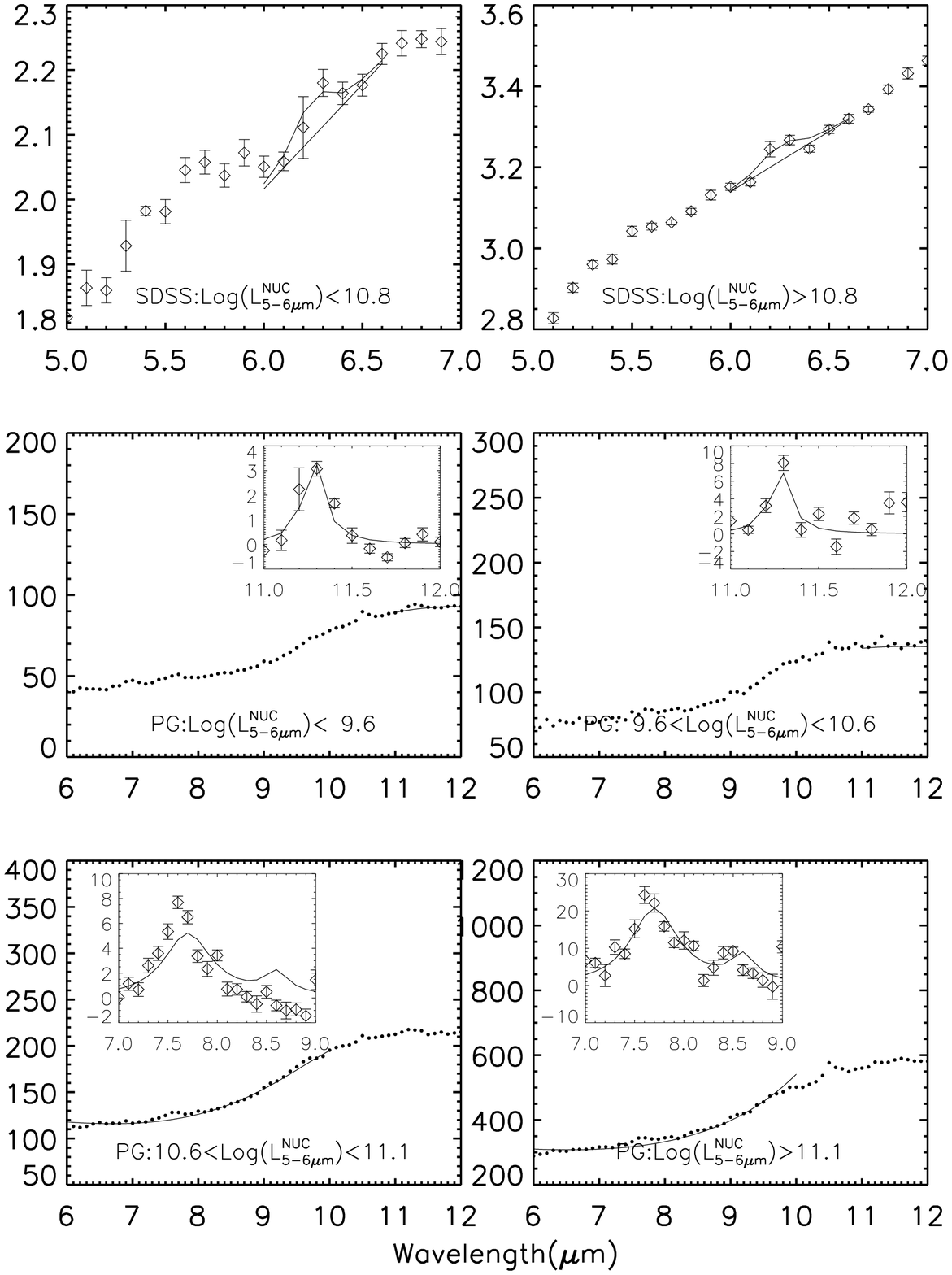}
\caption{\label{COMP_SPEC} The composite spectra of  PG and SDSS quasars without 
PAH-detection within given 5-6 $\mu$m luminosity ranges. }
\end{figure}

Figure~\ref{LSFIR_LQSO} shows the result for the whole PG sample of 90
objects (triangles) and  the whole SDSS sample of  57 objects (stars).
As  there is  no significant  redshift evolution  of the  SFRs  in the
quasar hosts  (\S~\ref{EVL-SFR}), the  two quasar samples  are plotted
together.    The   5-6  $\mu$m   IR   continuum  luminosity   ($L^{\rm
  NUC}_{5-6{\mu}m}$) is  used as a tracer of  nuclear luminosity. For clarity, the
uncertainty of the SFIR luminosity ($L_{\rm SFIR}$) is not shown in the figure.
A  significant number (50\%)  of objects  only have  upper-limits.  To
demonstrate  there  is  a  correlation  between  the  SFIR  luminosity
($L_{\rm SFIR}$) and nuclear luminosity, we stacked spectra of several
sub-samples  of  quasars without  aromatic  detections defined  within
certain  $L^{\rm  NUC}_{5-6{\mu}m}$ luminosity  bins.   The result  is
shown in Figure~\ref{COMP_SPEC}.  Each composite spectrum has detected
aromatic  features.  The  6.2 $\mu$m  feature is  measured as  in this
paper while  the 7.7 and  11.3 $\mu$m features are  measured following
\citet{Shi07}.     For    each    sub-sample,   the    mean    $L^{\rm
  NUC}_{5-6{\mu}m}$ is  used and the corresponding  SFIR luminosity is
measured  based on  the composite  spectrum.  The  result is  shown as
large    symbols     in    Figure~\ref{LSFIR_LQSO}.     The    $L^{\rm
  NUC}_{5-6{\mu}m}$  range   of  each   sub-sample  is  used   as  the
uncertainty  of the  $L^{\rm NUC}_{5-6{\mu}m}$  and  the corresponding
$L_{\rm SFIR}$ range is used as the uncertainty of the $L_{\rm SFIR}$.

To quantify  the correlation, linear  regression fits were  carried out
using  linmix$\_$err.pro that  accounts  for the  measured errors  and
upper-limits.   We performed fits  to three  data sets:  all individual
PG+SDSS data  points, all detected +  stacked data points  and all PG
objects. The  best-fit lines are shown  in Figure~\ref{LSFIR_LQSO} and
listed in  Table~\ref{Fit_LNUC_LSFIR}. As  shown in the  figure, the three
fits  give  roughly similar  results  and  have correlation coefficients  of
0.6-0.8.  Such  consistent  results   confirm  the  existence  of  the
correlation and  the lack of systematic difference  between $z=1$ SDSS
and low-redshift PG samples.

To  further demonstrate  that the  correlation  is not  caused by  the
aromatic feature non-detections in  about half the objects, we carried
out a  Monte-Carlo simulation for the  PG sample, which  spans a large
range  of SFIR luminosity.  Basically, a  SFIR luminosity  is assigned
randomly to  a PG quasar  with the relative probability  following the
SFIR  LF of  the  whole PG  sample.  In such  a  simulation, the  SFIR
luminosity is  assumed not to  correlate with the  nuclear luminosity.
For each  set of  simulations, a sub-sample  of objects  with aromatic
detections  can be  defined  through comparing  the simulated  $L_{\rm
  SFIR}$  and   the  observed  uncertainties   or  upper-limits.   The
Spearman's correlation coefficient can be measured for the Log($L_{\rm
  SFIR}$)$-$Log($L^{\rm  NUC}_{5-6{\mu}m}$)   of  such  a  sub-sample.
After 10000  simulations, the probability of  the simulated Spearman's
correlation coefficients  being higher than  the observed one  is only
1\%.

Both  of  these tests  imply  that  the  SFIR luminosity  most  likely
correlates with  the nuclear luminosity  in type-1 quasar  hosts.  The
comparison  of this  correlation  to those  in  theoretical models  is
presented in \S~\ref{DISSCUSS-M-sigma}.

\section{Discussion}\label{discussion}

\subsection{Comparison with Other Studies of AGNs at High-Redshift }

Intense star  formation in high-redshift quasars has  been observed in
some  individual quasars  as indicated  by rest-frame  far-IR emission
\citep[e.g.][]{Wang08},  aromatic  features \citep{Lutz08},  molecular
gas emission  \citep[for a  review, see][]{Solomon05} and  UV emission
\citep{Akiyama05}.   However,  these studies  suffer  from low  number
statistics, selection bias toward high SFRs and large uncertainties in
the derived SFR. \citet{Hatziminaoglou08} have studied IR SEDs of SDSS
quasars  in   the  SWIRE  field  through  fitting   torus  models  and
star-formation  templates.   They  derived on  average  star-formation
contributions of $\sim$35\% to the  total IR luminosity for quasars at
$z=$0.8-1.0, corresponding  to a  SFIR luminosity of  $\sim$ 10$^{12}$
L$_{\odot}$.  However, this  high SFR is only for  objects detected at
MIPS 70 and/or 160 $\mu$m, indicating a bias toward high SFRs in their
sample.   Our  study   of  a  complete  sample  of   57  SDSS  quasars
demonstrates conclusively that type-1  quasars reside in host galaxies
with  intense  star formation.  As  the  nucleus  is brighter  due  to
selection  effects  and the  host  galaxy  is  apparently smaller  and
fainter due  to surface-brightness dimming, the  host morphologies are
difficult   to   constrain    unambiguously   \citep[for   a   review,
  see][]{Davies08}  but  appear   to  be  consistent  with  elliptical
morphologies   \citep{Kukula01,  Hyvonen07,   Falomo04,  Kotilainen07,
  Falomo08,  Ammons09}.  We will  discuss this  apparent inconsistency
between  early-type  host morphology  and  intense  star formation  in
\S~\ref{DISCUSS-HOST}.

Similar to their low-redshift counterparts, IR-selected type 2 quasars
largely show disturbed morphologies \citep{Lacy07, Urrutia08} and have
intense  star formation  activity as  probed by  the  aromatic feature
\citep{Lacy07, Hernan-Caballero09}.  High SFRs  are also found for
optically-selected  type  2 quasars,  based  on  the aromatic  feature
\citep{Zakamska08}. High-$z$ ULIRGs with embedded AGNs (possibly related
to type 2 quasars) have also  been found to be associated with intense
star  formation  \citep[e.g.][]{Houck05,  Weedman06, Yan07,  Sajina07,
  Watabe09}.

A  correlation between the  mean stellar population age and the
AGN  luminosity   is  found  for   high-redshift  X-ray-selected  AGNs
\citep{Ammons09}.   Although  some   studies  do  not  confirm  strong
on-going  SFRs  in X-ray-selected  AGNs  with  a  range of  the  X-ray
luminosity  \citep{Sturm06,  Alonso-Herrero08,  Schawinski09}, a  more
thorough investigation of star  formation in a relatively large sample
with more plausible  probes should be made before  concluding that the
X-ray-selected  AGNs  are  in  a different  evolutionary  stage.   For
example, a  recent study  of X-ray-selected AGNs  in the  COSMOS field
indicates  enhanced SFRs  in  their host  galaxies  relative to  field
galaxies  with  similar stellar  masses  \citep{Silverman09}. It  also
suggested that a selection  bias may account for the X-ray-selected AGNs
 residing in the green valley region of the color-magnitude plot.

To  summarize,  AGNs  at   high  redshift  ($z$  $>$  0.5),  including
high-luminosity  type-1/type-2 quasars  and  relatively low-luminosity
X-ray-selected AGNs, are experiencing intense star-forming activity in
their host galaxies.

\subsection{Quasar Host Galaxies As A SFR-Enhanced Subset of Field Galaxies}\label{discussion1}

We have measured the aromatic-based  SFR for a complete sample of SDSS
type 1 quasars at $z\sim$ 1.  The derived SFIR LF is flatter than that
of $z\sim$ 1  field galaxies, implying enhanced SFRs  in the quasar hosts.
Combining  this with  studies  of IR-  and  optically-selected type  2
quasars  \citep{Lacy07, Zakamska08,  Hernan-Caballero09},  we conclude
that the  quasar host  galaxy population is  a SFR-enhanced  subset of
field  galaxies.  This  has been  known  to be  true for  low-redshift
quasars  as  shown  in  our  study  of  local  optically-selected  PG,
IR-selected 2MASS and radio-selected 3CR quasars \citep{Shi07}.

As discussed  in the  introduction, there are  many different  ways to
probe the stellar population age.  To interpret results with different
probes, one  must keep in  mind that each  probe is only  sensitive to
stellar  populations with  certain  ranges of  ages  and suffers  from
different  limitations.  Therefore,  apparently different results
about the stellar population age  of quasar hosts by different studies
are not  necessarily inconsistent with each other,  but instead should
be integrated to  achieve the complete view of  the stellar population
that  is critical  to understand  the BH/galaxy  co-evolution.  In the
following,  we reconcile our  result of  enhanced SFRs  with different
results in the literature.

\subsubsection{Comparisons with Host Galaxy Morphology Studies}\label{DISCUSS-HOST}

The IR-  and optically-selected type  2 quasar hosts show  high merger
($>$50\%)  fractions  at   both  low  \citep{Canalizo01,  Hutchings03,
  Marble03}    and    high    redshift    \citep{Lacy07,    Urrutia08,
  Hernan-Caballero09},   higher  than  field  galaxies of similar infrared luminosity
 \citep[$\sim$30\%; ][]{Shi09, Sobral09}.  The enhanced SFR  in type 2
quasars is thus consistent with their host morphologies.

For  optically/UV-selected  type 1  quasar  hosts, regular  early-type
morphologies  dominate, with  a  fraction of  $>$50\%  for quasars  at
$M_{B}$  $<$  -23  at  low  redshift  ($z$$<$0.5)  \citep{Hutchings84,
  Smith86, McLeod95, Bahcall97,  Dunlop03, Floyd04, Guyon06}.  At high
redshift, the  host galaxies appear  to be consistent  with elliptical
morphologies   \citep{Kukula01,  Hyvonen07,   Falomo04,  Kotilainen07,
  Falomo08,  Ammons09}, despite the  difficulties in  constraining the
types   unambiguously.    The   merger   fraction  is   $<$30\%,   not
significantly different from  field galaxies.  Therefore, the enhanced
SFR in  the type 1 quasar  hosts is not  a natural result of  the fact
that  the host  galaxy  is  massive, given  the  dominance of  regular
early-type  morphologies.

To reconcile  the apparent inconsistency  between elliptical-dominated
morphologies and enhanced SFRs, we propose that star formation in type
1 quasar  hosts occurs in the  circum-nuclear ($\sim$0.1-1kpc) region.
In  this   case,  any  star-forming  signature  would   be  missed  by
spatially-resolved  image  studies  that  always exclude  the  central
region as a  result of subtracting the diffracted  nuclear light.  The
large  extinction  in the  circumnuclear  region  can  also hide  star
formation traced by extinction-sensitive  probes (e.g.  UV emission \&
[OII]$\lambda$3727  lines)  \citep{Akiyama05,  Ho05, Kim06}.   Current
examples of direct detections  of intense circum-nuclear starbursts in
type-1 quasars  are limited to a few  nearby objects \citep{Cresci04}.
High-resolution  molecular  gas  mapping  has  revealed  a  large  gas
concentration  in  the  circum-nuclear  region  for  tens  of  quasars
\citep{Solomon05,   Maiolino07,   Riechers09,   Walter09}.    As   the
low-luminosity  counterparts  of   quasars,  nearby  Seyfert  galaxies
frequently  harbor   circum-nuclear  starbursts  \citep{Heckman97,  Gu01,
  GonzalezDelgado01,  CidFernandes04,  Davies07,  Riffel09},  although
enhanced star formation  also occurs in the spiral  disks of Seyfert 2
galaxies \citep{Maiolino95}.  Quasars, with higher luminosity and thus
larger  mass  inflow,  probably  harbor  more  intense  circum-nuclear
starbursts.  As shown in Figure~\ref{EVL_SF_QSO}(b), quasar hosts have
on  average LIRG-level SFRs.   In normal local  galaxies, such  intense star
formation  is  normally  achieved  only in  the  circumnuclear  region
\citep[for a review, see][]{Kennicutt98}.

\subsubsection{Comparisons with the Study of Stellar Populations}

Studies  of stellar  populations through  broad-band SEDs  and optical
spectra have discovered intermediate-age  stars or bluer colors in AGN
hosts  compared to  their  inactive galaxy  counterparts  at both  low
redshift   \citep{Ronnback96,  Brotherton99,   Kauffmann03,  Jahnke03,
  Jahnke04, Canalizo06, VandenBerk06, Jahnke07, Schawinski09} and high
redshift \citep{Ammons09}.   In the  following, we show  that the spectral
features employed  by current studies  of stellar populations  are not
sensitive to on-going ($<$0.1 Gyr) starburst activity.

The  continuum shape is  useful to  detect the  presence of  the young
stellar population only if the spectrum extends shorter than 4000$\AA$
or  the  broad-band SED  includes  U-band.   At  low redshift,  it  is
difficult to achieve  good coverage of the spectrum/SED  at such short
wavelengths.   Thus our result  of enhanced  SFRs does  not contradict
studies based on the continuum  shape at $\lambda$ $>$ 4000 $\AA$ that
detect   intermediate-age   stellar   populations   \citep{Ronnback96,
  Jahnke03, Jahnke04, Schawinski09, Ammons09}.   On the other hand, at
$\lambda$ $<$ 4000 $\AA$  the contrast between nuclear- and host-light
in type  1 AGNs becomes  much higher compared  to the value  at redder
wavelengths.  The  subtraction of the nuclear  light always introduces
large errors on  the resulting host brightness.  Even  in type 2 AGNs,
the  scattered  nuclear  UV  light  is  sometimes  important  or  even
dominates \citep{Zakamska06}.  More  importantly, obscuration can hide
blue host-galaxy light if  star formation occurs in the circum-nuclear
region  as  discussed  above.   Therefore, results  based  on  spectra
extending  shorter  than  $<$   4000  $\AA$  may  still  underestimate
significantly   the   level   of  on-going   star-formation   activity
\citep{Jahnke07, VandenBerk06, Chen09}.

\citet{Kauffmann03}  employed  the  4000  $\AA$  break  and  H$\delta$
absorption  to study  stellar populations  and  found intermediate-age
stars in hosts with early-type  morphologies.  The 4000 $\AA$ break is
caused by metal  line absorptions.  In hot stars with  age $<$ 0.1 Gyr
(see Fig.  2  of \citet{Kauffmann03b}), the 4000 $\AA$  break is small
($<$1.15) and insensitive to the stellar age. Moreover, the 4000 $\AA$
break is  a luminosity-weighted mean-age  indicator and thus  does not
have the temporal resolution to separate young stars from dominant old
stars in a massive host galaxy.  H$\delta$ absorption arises in late-B
to  early-F stars  and is  thus only  prominent 0.1-1  Gyr  after star
formation  \citep{Kauffmann03b}.  Therefore,  the two  probes  are not
sensitive  to  ongoing  star  formation activity.   In  addition,  
stellar-age-dependent obscuration may  also completely hide the effect
of massive  OB stars  on these two  features. In fact,  such selective
obscuration may  be the  reason for the  presence of some  LIRGs whose
optical  spectra  show  strong  H$\delta$ absorption  but  weak  O[II]
emission  \citep{Poggianti00}.   Studies   based  on  the  whole  SDSS
spectrum suffer  from similar problems  \citep{Jahnke07, VandenBerk06,
  Wild07,  Chen09},  as  the   continuum  shape  may  be  affected  by
extinction  and nuclear  light contamination  while all  the available
stellar  features (Balmer  break, high  order Balmer  absorption line,
4000$\AA$ break  and Ca  II (H\&K)) are  not sensitive to  the current
star formation whose  characteristic features (mainly nebular emission
lines) are contaminated severely by the nuclear radiation in both type
1 and type 2 objects.

In a summary, by integrating different studies,
we now have a more complete census of the stellar population in type-1
quasar hosts, which appear to be massive galaxies presumably dominated
by  old  stars, but  that  harbor  a  significant fraction  (10\%)  of
intermediate-age  stellar populations,  and  are experiencing  intense
circum-nuclear star formation.

\subsection{Implications For the BH-Bulge Correlation}\label{DISSCUSS-M-sigma}

\begin{figure}
\epsscale{1.2}
\plotone{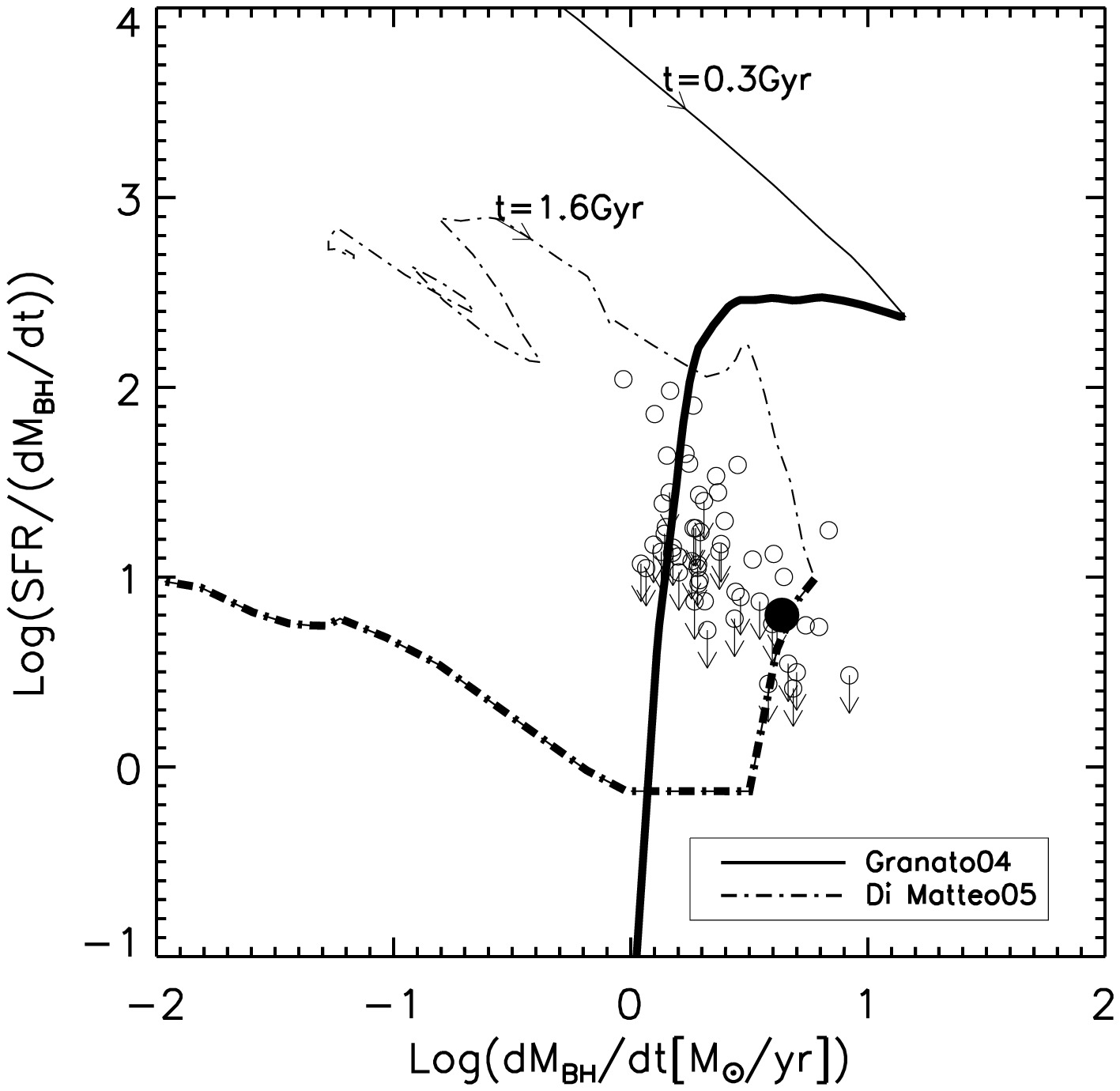}
\caption{\label{Mdot_SFR} The  ratio of the  SFR and the  BH accretion
rate vs. the BH accretion rate. The open circles show the results for
our $z\sim$1  quasars and the filled circle is the mean value derived
from  the  luminosity  function.   Curves  show these two quantities along the 
merging process in two models, where the solid line is for the model in \citet{Granato04} and
the dot-dashed line is for the model in \citet{DiMatteo05}.
The thick part of the curve indicates  the stage after
the peak BH accretion rate (presumably the type-1 quasar phase). }
\end{figure}

\begin{figure}
\epsscale{1.2}
\plotone{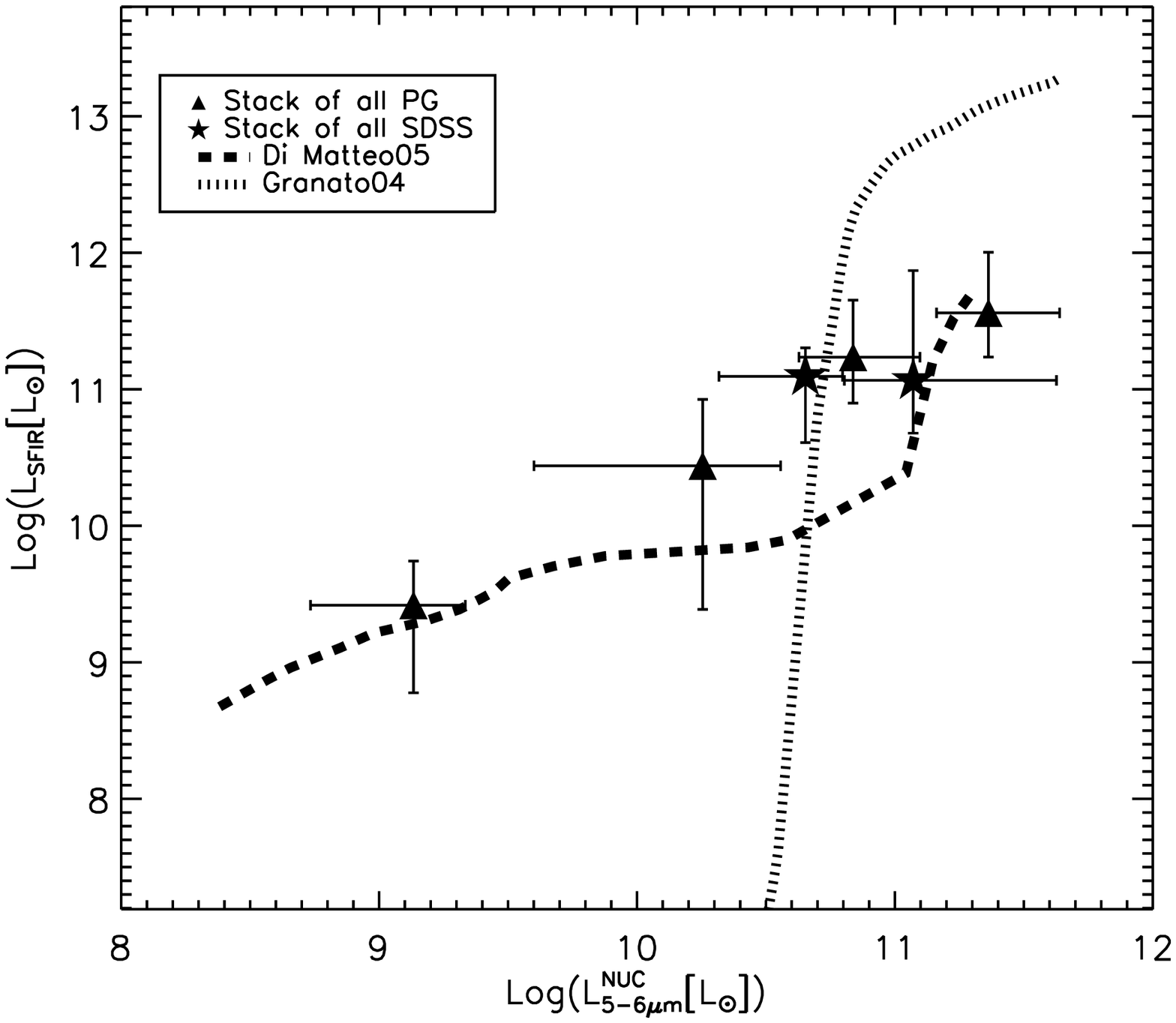}
\caption{ \label{LSFIR_LQSO_AVERAGE} The correlation between the star-forming IR luminosity and
the nuclear luminosity at 5-6 $\mu$m for the stacked spectra of all PG and SDSS quasars within certain
5-6 $\mu$m luminosity  ranges as indicated by their error-bars on the X-axis. These are compared to
the predictions of the model of \citet{DiMatteo05} and that of \citet{Granato04} 
shown as dashed and dotted lines, respectively.}
\end{figure}

The  direct measurements  of  SFRs  in quasar  hosts  can provide  new
insights  into the  mechanism  that shapes  the $M_{\rm  BH}$-$\sigma$
relationship. 
Figure~\ref{EVL_SF_QSO}(c) shows the  redshift evolution for the ratio
of the SFR in quasar hosts  to the black hole growth rate, obtained by
dividing  the integral  SFIR luminosity  function by  the  integral BH
growth rate  function.  The  SFIR luminosity is  converted to  the SFR
using  the \citet{Rieke09} relations  as shown  above.  The  BH growth
rate function is obtained from the quasar $B$-band luminosity function
using       $L_{\rm       bol}=\varepsilon/(1-\varepsilon)\dot{M}_{\rm
  BH}c^{2}=c_{B}L_{B}$,   where  $L_{\rm   bol}$  is   the  bolometric
luminosity, $\varepsilon$ is the mass to energy conversion efficiency,
$\dot{M}_{\rm BH}$ is the BH growth rate and $c_{B}$ is the bolometric
correction         for         the         $B$-band         luminosity
$L_{B}$.    Figure~\ref{EVL_SF_QSO}(c)    shows    the    result    for
$\varepsilon$=0.1   and  $c_{B}$=11.8,   where  the   quasar  $B$-band
luminosity    functions   are    given   by    \citet{Schmidt83}   and
\citet{Richards06}. As shown in the figure, the SFR/$\dot{M}_{\rm BH}$
in quasar  hosts is almost constant  with redshift and has  a value of
$\sim$10, much lower than  that expected from the $M_{\rm BH}$-$\sigma$
relationship. This result rules out the  simple model that
the  BH accretion and  star formation  evolve with  a fixed  ratio as
given by the $M_{\rm BH}$-$\sigma$ relation at any time for individual objects.
This implies that $M_{\rm BH}$/$M_{*}$ for type 1 quasars deviates from
that  implied by the  M$_{\rm BH}$-$\sigma$  relation.  The  factor of
deviation  of   $M_{\rm  BH}/M_{\rm  *}$  can  be   given  roughly  as
(1+${\Delta}M_{\rm       BH}/M_{\rm      BH}$)=exp$({\eta}\frac{t_{\rm
    qso}}{4.4{\times}10^{7}{\rm yr}})$, where $\eta$ is the fractional
Eddington  accretion   rate  $\eta$=$\dot{M}_{\rm  acc}$/$\dot{M}_{\rm
  acc}^{\rm  EDD}$ and  $t_{\rm qso}$  is the  duration of  the type-1
quasar   phase  in   years.   For   $\eta$  =   (0.1-1)   and  $t_{\rm
  qso}$=10$^{8}$ yr, the deviation is a factor of 1-10.

Numerical simulations have invoked  galaxy mergers and quasar feedback
to  successfully   explain  the  $M_{\rm   BH}$-$\sigma$  relationship
\citep{Granato04,  DiMatteo05}.   Different  models predict  different
time  evolution of the  SFR and  BH accretion  rate along  the merging
process. Our direct measurements of SFRs in type 1 quasars can provide
constrains on these models. Our $z\sim$1 quasars have a median BH mass
of      10$^{8.9\pm0.3}$      M$_{\odot}$     \citep{Shen08}.       In
Figure~\ref{Mdot_SFR},  we   compare  our  result  to   the  model  of
\citet{Granato04}  for their  most massive  dark matter  halo  and the
model of  \citet{DiMatteo05} for their most massive  galaxy.  For each
model, the  thicker part  of the curve  indicates the stage  after the
peak BH accretion rate, presumably  the type-1 quasar phase.  The open
circles  show the  results for  our  $z\sim$1 quasars  and the  filled
circle  is the  mean value  derived from  the luminosity  function. As
shown  in Figure~\ref{Mdot_SFR},  individual quasars  span a  range of
$\dot{M}_{\rm  BH}$  and SFR/$\dot{M}_{\rm  BH}$.   For  the model  of
\citet{Granato04},  the  full  range  of  the  SFR/$\dot{M}_{\rm  BH}$
resides within  their type  1 quasar phase.   However, the SFR  in the
model is more than an order of magnitude higher than our average value
(the  filled circle).   \citet{DiMatteo05} produced  a model  with the
mean $\dot{M}_{\rm  BH}$ and SFR/$\dot{M}_{\rm BH}$  comparable to our
observed  result. The  much  higher SFR  in  \citet{Granato04} may  be
related to their high star formation efficiency which is determined by
the  free-fall  or  cooling  time   in  their  models,  while  in  the
\citet{DiMatteo05} star formation occurs in a self-regulated mode.

In  Figure~\ref{LSFIR_LQSO_AVERAGE}, we  further  compare our  $L_{\rm
  SFIR}$  - $L^{\rm NUC}_{5-6{\mu}m}$  relationship to  predictions of
these two models. As models only produce the average case, we here use
the  $L_{\rm SFIR}$  - $L^{\rm  NUC}_{5-6{\mu}m}$  relation determined
from the composite  spectra of all PG and  SDSS quasars within certain
luminosity ranges.   The SFRs of the  two models are  converted to the
SFIR  luminosity  using  the  \citet{Rieke09}  relationship.   The  BH
accretion  rate is  first converted  to the  bolometric  luminosity by
assuming      $L_{\rm     bol}=\varepsilon/(1-\varepsilon)\dot{M}_{\rm
  BH}c^{2}$ and $\varepsilon$=0.1. The resulting bolometric luminosity
is  converted to the  $L^{\rm NUC}_{5-6{\mu}m}$  based on  the $L_{\rm
  bol}$-$L^{\rm  NUC}_{5-6{\mu}m}$  relationship  of our  SDSS  sample
whose bolometric  luminosity is from \citet{Shen08}.   Only the type-1
stage (i.e.,  after the peak BH  accretion rate) of the  two models is
shown    in    Figure~\ref{LSFIR_LQSO_AVERAGE}.     The    model    of
\citet{DiMatteo05}   is  generally   consistent   with  our   observed
relationship with a bit lower slope, while \citet{Granato04} predict a
much  steeper slope.   This implies  that the  quasar feedback  may be
overestimated significantly in \citet{Granato04}.  The feedback energy
in \citet{DiMatteo05} is  fixed to be 0.5\% of  the accreted rest mass
energy  while  this fraction  in  \citet{Granato04}  is several  times
higher.

\section{Conclusions}\label{conclusion}

To constrain  the cosmic  evolution of star  formation in  quasar host
galaxies, we present {\it Spitzer} IRS observations of a complete SDSS
sample of 57 type-1 quasars at $z\sim$1. The main conclusions are:

(1) About  half of the  sample has aromatic features  detected at
6.2 and/or 7.7 $\mu$m. The  composite profile of these two features is
similar to those  of normal galaxies, ULIRGs  and AGNs at both low and high redshift,
implying star-formation excitation of the aromatic features at $z\sim$1.

(2)  Based  on  the  aromatic-to-SFR  ratio  of  star-forming  galaxies  at
$z\sim$1, we  have constructed  the star-forming IR  (SFIR) luminosity
function  (LF)  of  $z\sim$1 type-1 quasars.   Similar  to  low-redshift  PG
quasars, these $z\sim$1  quasars show a flatter SFIR  LF than $z\sim$1
field  galaxies, implying  the quasar  host galaxy  population  has on
average higher SFRs.  Based on the measured SFIR LF, individual quasar
hosts are shown  to be experiencing on average  LIRG-level SFRs, which
most likely occur in the circumnuclear region.

(3) By comparing with similar measurements of low-redshift PG quasars,
we found that  the comoving SFIR energy density  in type-1 quasar hosts shows
much faster evolution  than that in field galaxies,  while the average
SFR and the average  SFR/$\dot{M}_{\rm acc}$ ratio are almost constant
with redshift.

(4) For  individual objects, we  have found a correlation  between the
aromatic-based  SFR and  the nuclear  luminosity.  By  comparing  this result  to
predictions  of different  models, we  have  found that  the model  in
\citet{Granato04} may over-estimate  the quasar feedback significantly
while the  model in  \citet{DiMatteo05} produces a  roughly consistent
result.

(5) By combining different studies in the literature, we now have a more
complete view of the stellar  population in type-1 quasar hosts, which reside
in  massive galaxies  with dominant  old stars,  harbor  a significant
fraction ($\sim$10\%) of  intermediate-age stellar populations and are
experiencing intense star formation.

\acknowledgements

We thank the anonymous referee for the detailed and constructive comments.
We also thank Jane Rigby and  Dean Hines for careful reading and comments.
Support for  this work was  provided by NASA through  contract 1255094
issued by JPL/California Institute of Technology.


\begin{thebibliography}{}

\bibitem[Akiyama(2005)]{Akiyama05} Akiyama, M.\ 2005, \apj, 629, 72 

\bibitem[Aller \& Richstone(2002)]{Aller02} Aller, M.~C., \& Richstone, D.\ 2002, \aj, 124, 3035 

\bibitem[Alonso-Herrero et al.(2008)]{Alonso-Herrero08} Alonso-Herrero, A., P{\'e}rez-Gonz{\'a}lez, P.~G., Rieke, G.~H., Alexander, D.~M., Rigby, 
J.~R., Papovich, C., Donley, J.~L., \& Rigopoulou, D.\ 2008, \apj, 677, 127 

\bibitem[Ammons et al.(2009)]{Ammons09} Ammons, S.~M., Melbourne, J., Max, C.~E., Koo, D.~C., \& Rosario, D.~J.~V.\ 2009, \aj, 137, 470 

\bibitem[Bahcall et al.(1997)]{Bahcall97} Bahcall, J.~N., Kirhakos, S., Saxe, D.~H., \& Schneider, D.~P.\ 1997, \apj, 479, 642 

\bibitem[Bennert et al.(2008)]{Bennert08} Bennert, N., Canalizo, G., Jungwiert, B., Stockton, A., Schweizer, F., Peng, C.~Y., \& Lacy, M.\ 2008, \apj, 677, 846 

\bibitem[Bertram et al.(2007)]{Bertram07} Bertram, T., Eckart, A., Fischer, S., Zuther, J., Straubmeier, C., Wisotzki, L., \& Krips, M.\ 2007, \aap, 470, 571 

\bibitem[Brand et al.(2008)]{Brand08} Brand, K., et al.\ 2008, \apj, 673, 119

\bibitem[Brandl et al.(2006)]{Brandl06} Brandl, B.~R., et al.\ 
2006, \apj, 653, 1129 

\bibitem[Brotherton et al.(1999)]{Brotherton99} Brotherton, M.~S., et al.\ 1999, \apjl, 520, L87 

\bibitem[Bruzual \& Charlot(2003)]{BC03} Bruzual, G., \& Charlot, S.\ 2003, \mnras, 344, 1000


\bibitem[Canalizo \& Stockton(2001)]{Canalizo01} Canalizo, G., \& Stockton, A.\ 2001, \apj, 555, 719 

\bibitem[Canalizo et al.(2006)]{Canalizo06} Canalizo, G., Stockton, A., Brotherton, M.~S., \& Lacy, M.\ 2006, New Astronomy Review, 50, 650 

\bibitem[Canalizo et al.(2007)]{Canalizo07} Canalizo, G., Bennert, N., Jungwiert, B., Stockton, A., Schweizer, F., Lacy, M., \& Peng, C.\ 2007, \apj, 669, 801 

\bibitem[Canalizo \& Stockton(2009)]{Canalizo09} Canalizo, G., \& Stockton, A.\ 2009, in preparation

\bibitem[Chen et al.(2009)]{Chen09} Chen, Y.-M., Wang, J.-M., Yan, C.-S., Hu, C., \& Zhang, S.\ 2009, \apjl, 695, L130 


\bibitem[Cid Fernandes et al.(2004)]{CidFernandes04} Cid Fernandes, 
R., Gu, Q., Melnick, J., Terlevich, E., Terlevich, R., Kunth, D., Rodrigues 
Lacerda, R., \& Joguet, B.\ 2004, \mnras, 355, 273 


\bibitem[Cresci et al.(2004)]{Cresci04} Cresci, G., Maiolino, R., Marconi, A., Mannucci, F., \& Granato, G.~L.\ 2004, \aap, 423, L13 


\bibitem[Cutri et al.(2002)]{Cutri02} Cutri, R.~M., Nelson, B.~O., Francis, P.~J., \& Smith, P.~S.\ 2002, IAU Colloq.~184: AGN Surveys, 284, 127 

\bibitem[Dale \& Helou(2002)]{Dale02} Dale, D.~A., \& Helou, G.\ 2002, \apj, 576, 159 




\bibitem[Davies et al.(2007)]{Davies07} Davies, R.~I., Mueller S{\'a}nchez, F., Genzel, R., Tacconi, L.~J., Hicks, E.~K.~S., Friedrich, 
S., \& Sternberg, A.\ 2007, \apj, 671, 1388 

\bibitem[Davies(2008)]{Davies08} Davies, R.\ 2008, New Astronomy Review, 52, 307 

\bibitem[Di Matteo et al.(2005)]{DiMatteo05} Di Matteo, T., Springel, V., \& Hernquist, L.\ 2005, \nat, 433, 604 

\bibitem[Dunlop et al.(2003)]{Dunlop03} Dunlop, J.~S., McLure, 
R.~J., Kukula, M.~J., Baum, S.~A., O'Dea, C.~P., 
\& Hughes, D.~H.\ 2003, \mnras, 340, 1095 

\bibitem[Engelbracht et al.(2008)]{Engelbracht08} Engelbracht, C.~W., 
Rieke, G.~H., Gordon, K.~D., Smith, J.-D.~T., Werner, M.~W., Moustakas, J., 
Willmer, C.~N.~A., \& Vanzi, L.\ 2008, \apj, 678, 804 

\bibitem[Evans et al.(2006)]{Evans06} Evans, A.~S., Solomon, P.~M., Tacconi, L.~J., Vavilkin, T., \& Downes, D.\ 2006, \aj, 132, 2398 


\bibitem[Falomo et al.(2004)]{Falomo04} Falomo, R., Kotilainen, 
J.~K., Pagani, C., Scarpa, R., \& Treves, A.\ 2004, \apj, 604, 495 

\bibitem[Falomo et al.(2008)]{Falomo08} Falomo, R., Treves, A., 
Kotilainen, J.~K., Scarpa, R., \& Uslenghi, M.\ 2008, \apj, 673, 694 

\bibitem[Farrah et al.(2008)]{Farrah08} Farrah, D., et al.\ 2008, \apj, 677, 957 


\bibitem[Ferrarese \& Merritt(2000)]{Ferrarese00} Ferrarese, L., \& Merritt, D.\ 2000, \apjl, 539, L9 

\bibitem[Floyd et al.(2004)]{Floyd04} Floyd, D.~J.~E., Kukula, 
M.~J., Dunlop, J.~S., McLure, R.~J., Miller, L., Percival, W.~J., Baum, 
S.~A., \& O'Dea, C.~P.\ 2004, \mnras, 355, 196 


\bibitem[Fu \& Stockton(2009)]{Fu09} Fu, H., \& Stockton, A.\ 2009, \apj, 696, 1693 

\bibitem[Gebhardt et al.(2000)]{Gebhardt00} Gebhardt, K., et al.\ 2000, \apjl, 539, L13 

\bibitem[Genzel et al.(1998)]{Genzel98} Genzel, R., et al.\ 1998, \apj, 498, 579 


\bibitem[Goldschmidt et al.(1992)]{Goldschmidt92} Goldschmidt, P., Miller, L., La Franca, F., \& Cristiani, S.\ 1992, \mnras, 256, 65P 

\bibitem[Gonz{\'a}lez Delgado et al.(2001)]{GonzalezDelgado01} Gonz{\'a}lez Delgado, R.~M., Heckman, T., \& Leitherer, C.\ 2001, \apj, 546, 845 

\bibitem[Granato et al.(2004)]{Granato04} Granato, G.~L., De Zotti, G., Silva, L., Bressan, A., \& Danese, L.\ 2004, \apj, 600, 580 

\bibitem[Gu et al.(2001)]{Gu01} Gu, Q.~S., Huang, J.~H., de Diego, J.~A., Dultzin-Hacyan, D., Lei, S.~J., \& Ben{\'{\i}}tez, E.\ 2001, \aap, 374, 932 

\bibitem[Guyon et al.(2006)]{Guyon06} Guyon, O., Sanders, D.~B., \& Stockton, A.\ 2006, \apjs, 166, 89 



\bibitem[Hatziminaoglou et al.(2008)]{Hatziminaoglou08} Hatziminaoglou, 
E., et al.\ 2008, \mnras, 386, 1252 


\bibitem[Heckman et al.(1997)]{Heckman97} Heckman, T.~M., Gonzalez-Delgado, R., Leitherer, C., Meurer, G.~R., Krolik, J., Wilson, 
A.~S., Koratkar, A., \& Kinney, A.\ 1997, \apj, 482, 114

\bibitem[Hern{\'a}n-Caballero et al.(2009)]{Hernan-Caballero09} Hern{\'a}n-Caballero, A., et al.\ 2009, \mnras, 395, 1695 


\bibitem[Ho(2005)]{Ho05} Ho, L.~C.\ 2005, \apj, 629, 680 



\bibitem[Houck et al.(2005)]{Houck05} Houck, J.~R., et al.\ 2005, \apjl, 622, L105 


\bibitem[Hughes et al.(2000)]{Hughes00} Hughes, D.~H., Kukula, M.~J., Dunlop, J.~S., \& Boroson, T.\ 2000, \mnras, 316, 204 

\bibitem[Hutchings et al.(1984)]{Hutchings84} Hutchings, J.~B., Crampton, D., \& Campbell, B.\ 1984, \apj, 280, 41 

\bibitem[Hutchings et al.(2003)]{Hutchings03} Hutchings, J.~B., 
Maddox, N., Cutri, R.~M., \& Nelson, B.~O.\ 2003, \aj, 126, 63 

\bibitem[Hyv{\"o}nen et al.(2007)]{Hyvonen07} Hyv{\"o}nen, T., Kotilainen, J.~K., {\"O}rndahl, E., Falomo, R., \& Uslenghi, M.\ 2007, \aap, 462, 525 

\bibitem[Jahnke \& Wisotzki(2003)]{Jahnke03} Jahnke, K., \& Wisotzki, L.\ 2003, \mnras, 346, 304 

\bibitem[Jahnke et al.(2004)]{Jahnke04} Jahnke, K., Kuhlbrodt, B., \& Wisotzki, L.\ 2004, \mnras, 352, 399 

\bibitem[Jahnke et al.(2007)]{Jahnke07} Jahnke, K., Wisotzki, L., Courbin, F., \& Letawe, G.\ 2007, \mnras, 378, 23 

\bibitem[Jester et al.(2005)]{Jester05} Jester, S., et al.\ 2005, \aj, 130, 873

\bibitem[Kauffmann et al.(2003)]{Kauffmann03} Kauffmann, G., et al.\ 2003, \mnras, 346, 1055 

\bibitem[Kauffmann et al.(2003b)]{Kauffmann03b} Kauffmann, G., et al.\ 2003, \mnras, 341, 33 

\bibitem[Kauffmann et al.(2007)]{Kauffmann07} Kauffmann, G., et al.\ 2007, \apjs, 173, 357 

\bibitem[Kelly(2007)]{Kelly07} Kelly, B.~C.\ 2007, \apj, 665, 1489 

\bibitem[Kennicutt(1998)]{Kennicutt98} Kennicutt, R.~C., Jr.\ 1998, \araa, 36, 189 

\bibitem[Kim et al.(2006)]{Kim06} Kim, M., Ho, L.~C., \& Im, M.\ 2006, \apj, 642, 702 

\bibitem[Kormendy \& Richstone(1995)]{Kormendy95} Kormendy, J., \& Richstone, D.\ 1995, \araa, 33, 581 

\bibitem[Kotilainen \& Ward(1994)]{Kotilainen94} Kotilainen, J.~K., \& Ward, M.~J.\ 1994, \mnras, 266, 953 

\bibitem[Kotilainen et al.(2007)]{Kotilainen07} Kotilainen, J.~K., Falomo, R., Labita, M., Treves, A., \& Uslenghi, M.\ 2007, \apj, 660, 1039 


\bibitem[Kukula et al.(2001)]{Kukula01} Kukula, M.~J., Dunlop, J.~S., McLure, R.~J., Miller, L., Percival, W.~J., Baum, S.~A., \& O'Dea, C.~P.\ 2001, \mnras, 326, 1533 

\bibitem[Lacy et al.(2007)]{Lacy07} Lacy, M., Sajina, A., Petric, A.~O., Seymour, N., Canalizo, G., Ridgway, S.~E., Armus, L., \& Storrie-Lombardi, L.~J.\ 2007, \apjl, 669, L61 


\bibitem[Le Floc'h et al.(2005)]{LeFloch05} Le Floc'h, E., et 
al.\ 2005, \apj, 632, 169 


\bibitem[Lutz et al.(2008)]{Lutz08} Lutz, D., et al.\ 2008, \apj, 684, 853 

\bibitem[Magnelli et al.(2009)]{Magnelli09} Magnelli, B., Elbaz, D., Chary, R.~R., Dickinson, M., Le Borgne, D., Frayer, D.~T., \& Willmer, C.~N.~A.\ 2009, \aap, 496, 57 

\bibitem[Magorrian et al.(1998)]{Magorrian98} Magorrian, J., et al.\ 1998, \aj, 115, 2285 


\bibitem[Maiolino et al.(1995)]{Maiolino95} Maiolino, R., Ruiz, M., Rieke, G.~H., \& Keller, L.~D.\ 1995, \apj, 446, 561 

\bibitem[Maiolino et al.(2007)]{Maiolino07} Maiolino, R., et al.\ 2007, \aap, 472, L33 

\bibitem[Marble et al.(2003)]{Marble03} Marble, A.~R., Hines, D.~C., Schmidt, G.~D., Smith, P.~S., Surace, J.~A., Armus, L., Cutri, 
R.~M., \& Nelson, B.~O.\ 2003, \apj, 590, 707 

\bibitem[Marconi et al.(2004)]{Marconi04} Marconi, A., Risaliti, G., Gilli, R., Hunt, L.~K., Maiolino, R., \& Salvati, M.\ 2004, \mnras, 351, 169 


\bibitem[McLeod \& Rieke(1995)]{McLeod95} McLeod, K.~K., \& Rieke, G.~H.\ 1995, \apjl, 454, L77 

\bibitem[Mickaelian et al.(2001)]{Mickaelian01} Mickaelian, A.~M., Gon{\c c}alves, A.~C., V{\'e}ron-Cetty, M.~P., \& V{\'e}ron, P.\ 2001, Astrophysics, 44, 14 

\bibitem[Murphy et al.(2009)]{Murphy09} Murphy, E.~J., Chary, 
R.-R., Alexander, D.~M., Dickinson, M., Magnelli, B., Morrison, G., Pope, 
A., \& Teplitz, H.~I.\ 2009, \apj, 698, 1380 


\bibitem[Netzer et al.(2007)]{Netzer07} Netzer, H., et al.\ 2007, \apj, 666, 806 

\bibitem[Nolan et al.(2001)]{Nolan01} Nolan, L.~A., Dunlop, J.~S., Kukula, M.~J., Hughes, D.~H., Boroson, T., \& Jimenez, R.\ 2001, \mnras, 323, 308 

\bibitem[Ogle et al.(2006)]{Ogle06} Ogle, P., Whysong, D., \& Antonucci, R.\ 2006, \apj, 647, 161 

\bibitem[Page \& Carrera(2000)]{Page00} Page, M.~J., \& Carrera, F.~J.\ 2000, \mnras, 311, 433 

\bibitem[P{\'e}rez-Gonz{\'a}lez et al.(2005)]{Perez-Gonzalez05} P{\'e}rez-Gonz{\'a}lez, P.~G., et al.\ 2005, \apj, 630, 82 


\bibitem[Poggianti \& Wu(2000)]{Poggianti00} Poggianti, B.~M., \& Wu, H.\ 2000, \apj, 529, 157 

\bibitem[Richards et al.(2001)]{Richards01} Richards, G.~T., et al.\ 2001, \aj, 121, 2308 

\bibitem[Richards et al.(2006)]{Richards06} Richards, G.~T., et 
al.\ 2006, \aj, 131, 2766 

\bibitem[Riechers et al.(2009)]{Riechers09} Riechers, D.~A., Walter, F., Carilli, C.~L., \& Lewis, G.~F.\ 2009, \apj, 690, 463 


\bibitem[Rieke et al.(2009)]{Rieke09} Rieke, G.~H., Alonso-Herrero, A., Weiner, B.~J., P{\'e}rez-Gonz{\'a}lez, P.~G., Blaylock, 
M., Donley, J.~L., \& Marcillac, D.\ 2009, \apj, 692, 556

\bibitem[Riffel et al.(2009)]{Riffel09} Riffel, R.~A., Storchi-Bergmann, T., Dors, O.~L., \& Winge, C.\ 2009, \mnras, 393, 783 



\bibitem[Rigby et al.(2008)]{Rigby08} Rigby, J.~R., et al.\ 2008, \apj, 675, 262 

\bibitem[Ronnback et al.(1996)]{Ronnback96} Ronnback, J., van Groningen, E., Wanders, I., \& \"Oumlrndahl, E.\ 1996, \mnras, 283, 282 

\bibitem[Roussel et al.(2001)]{Roussel01} Roussel, H., Sauvage, M., Vigroux, L., \& Bosma, A.\ 2001, \aap, 372, 427 


\bibitem[Sajina et al.(2007)]{Sajina07} Sajina, A., Yan, L., 
Armus, L., Choi, P., Fadda, D., Helou, G., 
\& Spoon, H.\ 2007, \apj, 664, 713 


\bibitem[Schawinski et al.(2009)]{Schawinski09} Schawinski, K., Virani, S., Simmons, B., Urry, C.~M., Treister, E., Kaviraj, S., 
\& Kushkuley, B.\ 2009, \apjl, 692, L19 

\bibitem[Schmidt \& Green(1983)]{Schmidt83} Schmidt, M., \& Green, R.~F.\ 1983, \apj, 269, 352 

\bibitem[Schweitzer et al.(2006)]{Schweitzer06} Schweitzer, M., et al.\ 2006, \apj, 649, 79 

\bibitem[Scoville et al.(2003)]{Scoville03} Scoville, N.~Z., Frayer, D.~T., Schinnerer, E., \& Christopher, M.\ 2003, \apjl, 585, L105 

\bibitem[Shankar et al.(2004)]{Shankar04} Shankar, F., Salucci, P., Granato, G.~L., De Zotti, G., \& Danese, L.\ 2004, \mnras, 354, 1020 

\bibitem[Shen et al.(2008)]{Shen08} Shen, Y., Greene, J.~E., Strauss, M.~A., Richards, G.~T., \& Schneider, D.~P.\ 2008, \apj, 680, 169 

\bibitem[Shi et al.(2006)]{Shi06} Shi, Y., et al.\ 2006, \apj, 653, 127 

\bibitem[Shi et al.(2007)]{Shi07} Shi, Y., et al.\ 2007, \apj, 669, 841 

\bibitem[Shi et al.(2009)]{Shi09} Shi, Y., Rieke, G., Lotz, J., \& Perez-Gonzalez, P.~G.\ 2009, \apj, 697, 1764 

\bibitem[Silverman et al.(2009)]{Silverman09} Silverman, J.~D., et al.\ 2009, \apj, 696, 396 


\bibitem[Smith et al.(1986)]{Smith86} Smith, E.~P., Heckman, 
T.~M., Bothun, G.~D., Romanishin, W., \& Balick, B.\ 1986, \apj, 306, 64 

\bibitem[Smith et al.(2007)]{Smith07} Smith, J.~D.~T., et al.\ 2007, \apj, 656, 770 

\bibitem[Sobral et al.(2009)]{Sobral09} Sobral, D., et al.\ 2009, \mnras, 949 

\bibitem[Solomon \& Vanden Bout(2005)]{Solomon05} Solomon, P.~M., \& Vanden Bout, P.~A.\ 2005, \araa, 43, 67

\bibitem[Soltan(1982)]{Soltan82} Soltan, A.\ 1982, \mnras, 200, 115 


\bibitem[Spoon et al.(2007)]{Spoon07} Spoon, H.~W.~W., 
Marshall, J.~A., Houck, J.~R., Elitzur, M., Hao, L., Armus, L., Brandl, 
B.~R., \& Charmandaris, V.\ 2007, \apjl, 654, L49 


\bibitem[Sturm et al.(2006)]{Sturm06} Sturm, E., Hasinger, G., Lehmann, I., Mainieri, V., Genzel, R., Lehnert, M.~D., Lutz, D., \& Tacconi, L.~J.\ 2006, \apj, 642, 81 


\bibitem[Teplitz et al.(2007)]{Teplitz07} Teplitz, H.~I., et al.\ 2007, \apj, 659, 941 

\bibitem[Urrutia et al.(2008)]{Urrutia08} Urrutia, T., Lacy, M., 
\& Becker, R.~H.\ 2008, \apj, 674, 80 

\bibitem[Valiante et al.(2007)]{Valiante07} Valiante, E., Lutz, D., Sturm, E., Genzel, R., Tacconi, L.~J., Lehnert, M.~D., \& Baker, A.~J.\ 2007, \apj, 660, 1060 

\bibitem[Vanden Berk et al.(2001)]{VandenBerk01} Vanden Berk, D.~E., et al.\ 2001, \aj, 122, 549 

\bibitem[Vanden Berk et al.(2006)]{VandenBerk06} Vanden Berk, D.~E., et al.\ 2006, \aj, 131, 84 


\bibitem[Walter et al.(2009)]{Walter09} Walter, F., Riechers, D., Cox, P., Neri, R., Carilli, C., Bertoldi, F., Weiss, A., \& Maiolino, R.\ 2009, \nat, 457, 699 


\bibitem[Watabe et al.(2009)]{Watabe09} Watabe, Y., Risaliti, 
G., Salvati, M., Nardini, E., Sani, E., 
\& Marconi, A.\ 2009, \mnras, 396, L1 


\bibitem[Wang et al.(2008)]{Wang08} Wang, R., et al.\ 2008, \apj, 687, 848 


\bibitem[Weedman et al.(2006)]{Weedman06} Weedman, D.~W., Le 
Floc'h, E., Higdon, S.~J.~U., Higdon, J.~L., 
\& Houck, J.~R.\ 2006, \apj, 638, 613 

\bibitem[Wild et al.(2007)]{Wild07} Wild, V., Kauffmann, G., 
Heckman, T., Charlot, S., Lemson, G., Brinchmann, J., Reichard, T., 
\& Pasquali, A.\ 2007, \mnras, 381, 543 


\bibitem[Wisotzki et al.(2000)]{Wisotzki00} Wisotzki, L., Christlieb, N., Bade, N., Beckmann, V., K{\"o}hler, T., Vanelle, C., \& Reimers, D.\ 2000, \aap, 358, 77 


\bibitem[Wu et al.(2005)]{Wu05} Wu, H., Cao, C., Hao, C.-N., Liu, F.-S., Wang, J.-L., Xia, X.-Y., Deng, Z.-G., \& Young, C.~K.-S.\ 2005, \apjl, 632, L79 

\bibitem[Yan et al.(2007)]{Yan07} Yan, L., et al.\ 2007, \apj, 658, 778 

\bibitem[Yu \& Tremaine(2002)]{Yu02} Yu, Q., \& Tremaine, S.\ 2002, \mnras, 335, 965 

\bibitem[Zakamska et al.(2006)]{Zakamska06} Zakamska, N.~L., et 
al.\ 2006, \aj, 132, 1496 

\bibitem[Zakamska et al.(2008)]{Zakamska08} Zakamska, N.~L., 
G{\'o}mez, L., Strauss, M.~A., \& Krolik, J.~H.\ 2008, \aj, 136, 1607 

\end{thebibliography}
\end{document}